\documentclass[11pt]{article}
\usepackage{fancyheadings}
\usepackage{amssymb,amsmath}
\usepackage{ascmac}
\usepackage[dvipdfmx]{hyperref}
\usepackage[dvips]{graphicx,psfrag}
\usepackage{epsfig}
\usepackage{array}
\usepackage{lscape}
\usepackage{accents}
\usepackage{color}
\usepackage{braket}
\usepackage{bm}
\usepackage{multirow}

\numberwithin{equation}{section}

\makeatletter
\def\alt{\mathrel{\mathpalette\gl@align<}}
\def\agt{\mathrel{\mathpalette\gl@align>}}
\def\gl@align#1#2{\lower.6ex\vbox{\baselineskip\z@skip\lineskip\z@
\ialign{$\m@th#1\hfil##\hfil$\crcr#2\crcr\sim\crcr}}}
\makeatother

\newcommand{\doi}[1]{\href{https://doi.org/#1}{doi:#1}}

\newcommand{\dash}{\hspace{1pt}--\hspace{1pt}}

\begin{document}

\begin{center}
\baselineskip 20pt

%%% TITLE %%%
{\bf
%\fontsize{16pt}{0pt}\selectfont
\LARGE
Insights from the magnetic field dependence of
the muonium-to-antimuonium transition 
} 

\vspace{1.0cm}

%%% AUTHORS %%%
{\Large 
Takeshi Fukuyama$\,{}^{a}$,
Yukihiro Mimura$\,{}^{b}$
and
Yuichi Uesaka$\,{}^{c}$
}

\vspace{8mm}

%%% AFFILIATION %%%
{\large
${}^a${\it 
Research Center for Nuclear Physics (RCNP),
Osaka University, \\Ibaraki, Osaka, 567-0047, Japan
}\\
\vspace{3mm}
${}^{b}${\it
Department of Physical Sciences, College of Science and Engineering, \\
Ritsumeikan University, Shiga 525-8577, Japan
}\\
\vspace{3mm}
${}^{c}${\it
Faculty of Science and Engineering, Kyushu Sangyo University, \\
2-3-1 Matsukadai, Higashi-ku, Fukuoka 813-8503, Japan
}\\
}

\vspace{1.2cm}

{\Large
{\bf Abstract}}\end{center}
%\vspace{3mm}
\baselineskip 18pt
{\large
The muonium-to-antimuonium transition experiment is about to be updated. 
Notably, the experiment at J-PARC in Japan can explore the magnetic field dependence
of the transition probability.
In this paper, we investigate the information that we can extract from 
the transition probabilities across different magnetic field strengths,
while also taking into account 
a planned transition experiment at CSNS in China. 
There are two model-independent parameters in the transition amplitude,
and we ascertain the feasibility of determining these parameters, including their relative physical phase,
from experimental measurements.
This physical phase can be related to the electron electric dipole moment, 
 which is severely constrained by experiments. 
The underlying mediator responsible for the transition
can be either doubly charged particles or neutral particles. 
In the former case, typical magnetic fields yield specific probability ratios, 
while the latter presents a range of the probability ratio.
We investigate several models with neutral mediators, and elucidate that 
the probability ratio is linked to the sign
of new physics contribution to the electron $g-2$.
The pivotal role of the J-PARC transition experiment in shedding light on these insights is emphasized.
}

\thispagestyle{empty}
\newpage
\addtocounter{page}{-1}

\section{Introduction}
\label{sec1}

\fontsize{13pt}{18pt}\selectfont

High-intensity muon beamlines are undergoing upgrades \cite{Kawamura:2018apy,Aiba:2021bxe},
opening the door to various studies on muon material physics.
Among these endeavors, the exploration of lepton flavor violation (LFV)
has garnered significant attention, encompassing processes such as
 $\mu \to e\gamma$ \cite{Adam:2013mnn},
$\mu \to 3 e$ \cite{Bellgardt:1987du}, and $\mu \to e$ conversion in nuclei \cite{SINDRUMII:2006dvw}.
These investigations are particularly significant as they delve into physics beyond the standard model (SM).
The muon facilities will also investigate the
transition of muonium ($\mu^+ e^-$)
into antimuonium ($\mu^- e^+$) (Mu-to-$\overline{\rm Mu}$ transition) \cite{Pontecorvo:1957cp,Feinberg:1961zza,Lee:1977tib,Halprin:1982wm}.
While a quarter-century has passed since
the transition experiment at Paul Scherrer Institute (PSI) set the most stringent constraint \cite{Willmann:1998gd},
the upcoming transition experiments
are ready to reinvigorate this pursuit,
as exemplified by the
Muonium-to-Antimuonium Conversion Experiment (MACE) at China Spallation Neutron Source (CSNS) \cite{Han:2021nod,Bai:2022sxq}
and an experiment using a brand-new approach at Japan Proton Accelerator Research Complex (J-PARC) 
 \cite{Kawamura:2021lqk}.
In the past twenty-five years, 
our understanding of the lepton sector has been improved 
by other experiments such as neutrino oscillations,
thus developing the theoretical environment for the Mu-to-$\overline{\rm Mu}$ transition
\cite{Conlin:2020veq,Fukuyama:2021iyw}.

Global non-abelian flavor symmetries in gauge interactions with quarks and leptons
suppress the occurrence of flavor changing neutral currents (FCNCs).
Mass differences among quarks and leptons violate the global flavor symmetries.
Therefore, FCNCs are radiatively induced typically in the down-type quark sector.
In the SM, FCNCs are negligible in the charged lepton sector, 
owing to the minuscule mass differences of neutrinos. 
If there is a new particle beyond the SM,
new couplings between the new particle and the leptons can serve as new sources of LFV that are potentially detectable by experiments.
Therefore, searching charged lepton flavor changes can be an effective tool for probing new physics beyond the SM. 
The absence of LFV decays 
requires specific flavor symmetries or parameter arrangements.
For instance, introducing an additional Higgs doublet that couples to fermions usually involves 
the selection of which doublet can couple to generate the masses of up- and down-type quarks as well as charged leptons.
Alternatively, nearly aligned Yukawa coupling matrices are required to suppress FCNCs.
In the lepton sector, we can assume a discrete flavor symmetry 
to eliminate muon flavor violating decays, which are severely constrained by experiments.
Even if $\Delta L_e = - \Delta L_\mu = \pm 1$ processes are prohibited,
$\Delta L_e = - \Delta L_\mu = \pm 2$ processes may still be allowed,
resulting in the generation of 
the Mu-to-$\overline{\rm Mu}$ transition at the tree level.
Neutral or doubly charged particles could serve as 
mediators of the $\Delta L_e = - \Delta L_\mu = \pm 2$ process.

The PSI experiment which provides the current bound on the Mu-to-$\overline{\rm Mu}$ transition 
attempts to detect electrons from the decays of $\mu^-$ in $\overline{\rm Mu}$ that is expected from the 
transition
in the presence of a magnetic field.
Because the experiment cannot specify the decay time elapsed since Mu production,
it obtains a bound of the time-integrated transition probability.
We refer to this measurement technique as the PSI method.
The upcoming MACE experiment in China will adopt the PSI method for measurement.
In contrast, 
the J-PARC experiment 
will measure the time-dependent probability of the transition.
At a specific time, a laser ionizes $\overline{\rm Mu}$ that is expected from the transition,
and the resulting dissolved $\mu^-$ is then carried by an electric field and directed towards a spectrometer.
We refer to this measurement technique as the J-PARC method.
The J-PARC method allows for alteration of
the magnetic field where Mu is produced,
and enables, in principle, the measurement of the magnetic field dependence of the transition probability.
The dependence of the transition probability on the magnetic field
varies depending on the operators to induce the transition \cite{Horikawa:1995ae,Hou:1995np}.
To distinguish the mediator of the transition,
it is crucial to measure the transition probabilities under different magnetic fields.
If doubly charged particles act as mediators,
the operators induced by them yield specific ratios of the transition probabilities.
Therefore, measuring probability ratios can quickly identify the operators.
On the other hand, if neutral particles are mediators, the ratios can take on various values 
depending on model parameters. Therefore, closer analyses of models are necessary.

In this paper,
we study the magnetic field dependence of the transition probabilities induced by neutral mediators,
which may be found by the combination of the J-PARC and MACE experiments.
The J-PARC method can measure 
 the transition probabilities both at
a weak magnetic field $B \alt 1 \, \mu$T
and at a medium magnetic field approximately equal to the geomagnetic field $\sim O(10) \,\mu$T.
The MACE experiment (and possibly an upgraded experiment at PSI) 
will measure the time-integrated probability at $B = 0.1\,$T.
Two parameters exist for the model-independent description of the Mu-to-$\overline{\rm Mu}$ transition amplitudes.
The transition probabilities at three magnetic fields,
together with information on the muon polarization in the produced Mu at $B=0.1\,$T,
enable the determination of the two parameters with a possible relative phase in the amplitudes. 
When the transition is induced by a single mediator,
the relative phase is related to the electric dipole moment (EDM) of the electron,
and thus, the phase should be very small due to the experimental bound on the electron EDM \cite{ACME:2018yjb,Roussy:2022cmp}.
In this study, we will examine three possible neutral mediators that can induce the transition: 
(1) Axion-like particle (ALP), (2) Inert doublet model, and (3) Neutral flavor gauge boson.
We assume that the models mentioned above do not induce the 
 $\Delta L_e = - \Delta L_\mu = \pm 1$ processes.
Even under the assumption, the electron and muon masses as well as their anomalous magnetic moments ($g-2$)
can be modified radiatively.
These models can make a significant contribution to the electron $g-2$ due to the flavor violation,
resulting from the chirality flip caused by the muon mass at the internal line of the loop diagram
for the electron $g-2$.
The current bound of the Mu-to-$\overline{\rm Mu}$ transition
restricts the contributions to the muon and electron $g-2$.
The new physics contribution to the electron $g-2$ ($\Delta a_e$) can be either positive or negative
(see Eqs.(\ref{eq:e-gminus2-Cs}) and (\ref{eq:e-gminus2-Rb}) for current status 
of the electron $g-2$).
For the scalar mediators (1) and (2), 
the transition bound allows for a significant value of $|\Delta a_e|$.
We emphasize that
the magnetic field dependence of the transition probability is linked
to the sign of $\Delta a_e$, 
and the measurements at J-PARC can impact these models.

This paper is organized as follows:
In Section \ref{sec2}, we review Mu-to-$\overline{\rm Mu}$ transition operators
and the transition probability as a function of the operator coefficients 
and magnetic field in the presence of non-relativistic Mu.
In Section \ref{sec3},
we explore the magnetic field dependence of the amplitude
in each transition operator.
In Section \ref{sec4},
we define the ratios of the transition probabilities
at three magnetic fields in the J-PARC and PSI methods,
and analyze what we can deduce from the ratios.
In Section \ref{sec5},
we study the model with ALP
and the relationship between the electron $g-2$ and the magnetic field dependence of the
transition probability.
In Section \ref{sec6},
we study the inert doublet model
and describe the muon and electron $g-2$ in the model.
By examining the ratio of the transition probability, 
it is possible to investigate the parameters of the model and its consistency with the electron $g-2$.
In Section \ref{sec7},
we describe the ratio of the transition probability in the model with neutral flavor gauge boson.
Additionally, we mention the relationship between a muon decay parameter 
and the magnetic field dependence of the transition probability in this model.
Section \ref{sec8} is dedicated to the conclusion.
In Appendix \ref{appendix:A}, we overview the energy eigenstates of Mu and $\overline{\rm Mu}$ in a magnetic field.
In Appendix \ref{appendix:B}, the populations of the states in the produced Mu are described.
In Appendix \ref{appendix:C}, we revisit the transition amplitudes in non-relativistic states
to help understand the magnetic field dependence of the amplitude in each operator.
We provide an explanation for the presence of two model-independent parameters 
in the transition amplitudes, 
despite there being five independent operators.

\section{Brief review of the Mu-to-$\overline{\bf Mu}$ transition probability}
\label{sec2}

This section reviews the probability
of the Mu-to-$\overline{\rm Mu}$ transition
and its magnetic field dependence \cite{Horikawa:1995ae,Hou:1995np}.
Appendix \ref{appendix:A} describes 
the four states of the Mu ground state that arise from combining the spins of $\mu^+$ and $e^-$: 
${\bf 2} \times {\bf 2} = {\bf 3}+ {\bf 1}$.
These states can be labeled by quantum numbers $(F,m)$,
where $F$ denotes the magnitude of total angular momentum
and $m$ signifies the $z$-component of total angular momentum.
The $F=1$ (triplet) and $F=0$ (singlet) states can exhibit distinct transition amplitudes
depending on operators that induce the Mu-to-$\overline{\rm Mu}$ transition.
How these states respond in a magnetic field hinges on their quantum numbers. 
Consequently, the magnetic field dependence on the transition probability 
assists in discerning the class of operators.

The four-fermion operators of the Mu-to-$\overline{\rm Mu}$ transitions are given as \cite{Conlin:2020veq}
\begin{align}
Q_1 &= (\bar \mu \gamma_\alpha (1-\gamma_5) e) (\bar \mu \gamma^\alpha (1-\gamma_5) e), \label{eq:Q1}\\
Q_2 &= (\bar \mu \gamma_\alpha (1+\gamma_5) e) (\bar \mu \gamma^\alpha (1+\gamma_5) e), \label{eq:Q2}\\
Q_3 &= (\bar \mu \gamma_\alpha (1+\gamma_5) e) (\bar \mu \gamma^\alpha (1-\gamma_5) e), \label{eq:Q3}\\
Q_4 &= (\bar \mu  (1-\gamma_5) e) (\bar \mu  (1-\gamma_5) e), \label{eq:Q4}\\
Q_5 &= (\bar \mu  (1+\gamma_5) e) (\bar \mu  (1+\gamma_5) e). \label{eq:Q5}
\end{align}
Any four-fermion operators of the transitions 
can be expressed as a linear combination of the five operators utilizing Fierz transformation.
For example, $S\times S$ and $P \times P$ operators are given as follows:
\begin{align}
Q_S &= (\bar \mu e) (\bar \mu e) = \frac14 ( -Q_3+ Q_4 + Q_5) , \label{eq:QS} \\
Q_P & = (\bar \mu \gamma_5 e) (\bar \mu \gamma_5 e) = \frac14 ( Q_3+ Q_4 + Q_5) \label{eq:QP}. 
\end{align}
When the Hamiltonian for the Mu-to-$\overline{\rm Mu}$ transition is given as
\begin{equation}
{\cal H}_{\rm Mu\mbox{-}to\mbox{-}\overline{Mu}} = \frac1{\sqrt2} (G_1 Q_1 + G_2 Q_2 + G_3 Q_3 + G_4 Q_4 + G_5 Q_5),
\end{equation}
the transition amplitudes ${\cal M}_{F,m} \equiv \langle \overline{\rm Mu}; F,m| {\cal H}_{\rm Mu\mbox{-}to\mbox{-}
\overline{Mu}} | {\rm Mu} ; F,m\rangle  $ for the four states, $(F,m) = (1,\pm 1), (1,0), (0,0)$,
in a non-relativistic limit are
obtained as (see Ref.\cite{Fukuyama:2021iyw})
\begin{align}
{\cal M}_{1,m} &= - \frac{8 |\varphi(0)|^2}{\sqrt2} \left(G_0 +  \frac12 G_3 \right),  \label{eq:M1m}\\
{\cal M}_{0,0} &= - \frac{8 |\varphi(0)|^2}{\sqrt2} \left(G_0 - \frac32 G_3 \right), \label{eq:M00}
\end{align}
where we define
\begin{equation}
G_0 \equiv G_1 + G_2 -\frac14 G_4 - \frac14 G_5.
\end{equation}
The wave function of an electron at the position of a muon is given by
\begin{equation}
|\varphi (0)|^2 = \frac{(m_{\rm red} \alpha)^3}{\pi} , \qquad
m_{\rm red} = \frac{m_\mu m_e}{m_\mu + m_e} ,
\end{equation}
where $\alpha$ is the fine structure constant.

The Mu-to-$\overline{\rm Mu}$ transition probability at time $t$ is expressed as
\begin{equation}
P(t) = \sum_{(F,m)} f_{F,m} P(F,m; t),
\label{Pt}
\end{equation}
where the coefficients $f_{F,m}$ correspond to the diagonal elements of the density matrix 
for the states $|{\rm Mu}; F,m\rangle_B$,
and represent the population of each state of the produced Mu.
The subscript $B$ attached to the states indicates that these states are energy eigenstates in a magnetic field.

The transition probabilities of the $m = \pm 1$ %$|1,\pm 1\rangle_B$ 
states
are given as
\begin{equation}
P(1,\pm1; t) \simeq e^{-\Gamma t} \frac{|{\cal M}_{1,\pm 1}|^2}{|{\cal M}_{1,\pm 1}|^2 + (\Delta E/2)^2}
\sin^2 \sqrt{|{\cal M}_{1,\pm 1}|^2 + (\Delta E/2)^2} t,
\end{equation}
where $\Gamma$ is the decay width of Mu, and $\Delta E$  %= aY$ 
is the energy splitting between $|{\rm Mu};1,\pm1 \rangle_B$ and $|\overline{\rm Mu};1,\pm 1\rangle_B$ 
in the presence of a magnetic field.
For the $m=0$ states, the oscillation time is much longer than the lifetime $\tau = 1/\Gamma \simeq 2.2\,\mu$s, and
the probabilities are approximately given as
\begin{equation}
P(F,0;t) \simeq e^{-\Gamma t} |{\cal M}_{F,0}^B|^2 t^2.
\end{equation}
The $m=0$ states are mixed
in the presence of a magnetic field, and the transition amplitudes 
are given as (see Appendix \ref{appendix:A} for their mixing in the presence of the magnetic field $B$)
\begin{align}
{\cal M}_{1,0}^B &= C^2 {\cal M}_{1,0} - S^2 {\cal M}_{0,0} = \frac{{\cal M}_{1,0} - {\cal M}_{0,0}}{2} + \frac{{\cal M}_{1,0} + {\cal M}_{0,0}}{2\sqrt{1+X^2}}  \nonumber \\
&
= - \frac{8 |\varphi(0)|^2}{\sqrt2} \left( G_3 + \frac{G_0 - \frac12 G_3}{\sqrt{1+X^2}}  \right), \label{eq:M10B} \\
{\cal M}_{0,0}^B &= C^2 {\cal M}_{0,0} - S^2 {\cal M}_{1,0} = \frac{{\cal M}_{0,0} - {\cal M}_{1,0}}{2} + \frac{{\cal M}_{1,0} + {\cal M}_{0,0}}{2\sqrt{1+X^2}}  \nonumber \\
&
= - \frac{8 |\varphi(0)|^2}{\sqrt2} \left( -G_3 + \frac{G_0 - \frac12 G_3}{\sqrt{1+X^2}} \right), \label{eq:M00B}
\end{align}
where $X$ is defined in Eq.(\ref{eq:X}):
\begin{equation}
X  \simeq 6.31 \times \frac{B}{\rm Tesla}.
\end{equation}
We assume that the coefficients $f_{F,m}$ satisfy
\begin{equation}
f_{1,1} + f_{1,-1} = f_{1,0} + f_{0,0} = \frac12.
\end{equation}
Refer to Appendix \ref{appendix:B} for more detailed information.
The total transition probability is then expressed as
\begin{align}
P(t) \simeq e^{-\Gamma t} 
&
 \Bigl( f_{1,0} |{\cal M}_{1,0}^B|^2 t^2+ f_{0,0} |{\cal M}_{0,0}^B|^2 t^2  
 \nonumber\\
& 
+ \frac12
\frac{|{\cal M}_{1,1}|^2}{|{\cal M}_{1, 1}|^2  + (\Delta E/2)^2} 
\left.
 \sin^2 \sqrt{|{\cal M}_{1, 1}|^2 + (\Delta E/2)^2} t \, \right).
\label{time-dependent-prob}
\end{align}
The time-integrated transition probability is calculated as
\begin{align}
\bar P &= \int_0^\infty \Gamma P(t) dt
\simeq 
2\tau^2 \left( f_{1,0} |{\cal M}_{1,0}^B|^2 + f_{0,0} |{\cal M}_{0,0}^B|^2 + \frac12 \frac{|{\cal M}_{1,1}|^2}{1+ (\tau \Delta E)^2} \right).
\label{time-integrated-prob}
\end{align}
We here note the following numerical values: 
\begin{align}
&\tau \Delta E = 3.85 \times 10^5 \times \frac{B}{\rm Tesla}, \\
&(8|\varphi(0)|^2)^2 \tau^2 G_F^2
= \frac{64 m_{\rm red}^2 \alpha^6 \tau^2 G_F^2}{\pi^2} = 2.57 \times 10^{-5}.
\end{align}
As described in Appendix \ref{appendix:B},
we assume that the coefficients $f_{F,0}$ are given by
\begin{equation}
f_{1,0} = \frac14 \left( 1 - P_\mu \frac{X}{\sqrt{1+X^2}} \right), \qquad
f_{0,0} = \frac14 \left( 1 + P_\mu \frac{X}{\sqrt{1+X^2}} \right).
\label{eq:f10-f00}
\end{equation}
The parameter $P_\mu$ is defined by Eq.(\ref{eq:Pmu}),
and it can be interpreted as the muon polarization in the produced Mu.
For $B \alt 1 \, {\rm mT}$, 
 the coefficients can be approximated as
\begin{equation}
f_{1,0} \simeq f_{0,0} \simeq \frac14.
\end{equation}

When the magnetic field is very weak, i.e., $B = B_0 \ll 1\, \mu$T,
the $m= \pm 1$ states can fully contribute to the Mu-to-$\overline{\rm Mu}$ transition.
We obtain 
\begin{align}
P(t, B=B_0) &\simeq 1.3 \times 10^{-5} \times  e^{-\Gamma t} \frac{t^2}{\tau^2}  
\left(   \left|  \frac{G_0}{G_F} \right|^2 
+ 
\frac34 \left|  \frac{ G_3}{G_F} \right|^2 \right)  \label{time-dependent-B1}. 
\end{align}
When the magnetic filed is $B \agt O(10) \ \mu$T, the contribution 
of the $m = \pm 1$
states can be dropped,
and the probability is 
\begin{align}
P(t) &\simeq 1.3 \times 10^{-5} \times  e^{-\Gamma t} \frac{t^2}{\tau^2}  
\left(f_{1,0}  \left| \frac{G_3}{G_F} + \frac{G_0 - \frac12 G_3}{G_F\sqrt{1+X^2}} \right|^2 
+ 
f_{0,0} \left| - \frac{G_3}{G_F} + \frac{G_0 - \frac12 G_3}{G_F\sqrt{1+X^2}} \right|^2 \right) .
  \label{time-dependent-B2}
\end{align}
We note $\int_0^\infty \Gamma e^{-\Gamma t} t^2 dt = 2\tau^2$ to obtain the time-integrated probability $\bar P$.
If 
$G_3 = 0$,
one simply obtains
\begin{equation}
\bar P 
=2.6 \times 10^{-5} \times 
\left| \frac{G_0}{G_F} \right |^2 \frac1{2(1+X^2)}.
\end{equation}
The PSI experiment has obtained the bound on the time-integrated probability under $B = 0.1\,$T \cite{Willmann:1998gd},
\begin{equation}
\bar P < 8.3 \times 10^{-11},
\end{equation}
which is translated to
\begin{equation}
\frac{|G_0|}{G_F} < 3.0 \times 10^{-3}.
\end{equation}

\begin{figure}[t]
\center
\includegraphics[width=12cm]{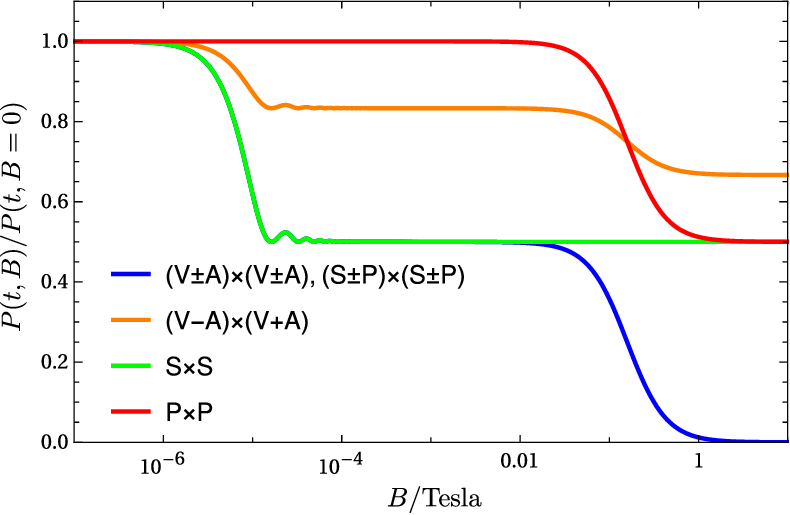}
\caption{
Magnetic field dependence of the ratios of time-dependent probabilities to those at $B=0$,
corresponding to various transition operators.
}
\label{fig1}
\end{figure}

Fig.\ref{fig1} illustrates 
the ratio of the transition probability under magnetic field $B$
to the probability when there is no magnetic field.
The muon polarization in the produced Mu is assumed to be $P_\mu = 0$ for simplicity.
A time, $t = \tau = 2.2\, \mu$s, is chosen for plotting the figure.
The suppression of the transition of the $m= \pm 1$ states
slightly depends on the chosen time.
For $B \agt O(10) \, \mu$T, the transition of the $m= \pm 1$ states
is suppressed, and the ratio of transition probabilities does not depend on the choice,
resulting in consistent ratios of the time-integrated probabilities $\bar P$ in the magnetic field.
The blue line (overlapping with the green line for $B \alt 0.01\,$T) is the plot 
for $(V\pm A)\times (V\pm A)$ and $(S\pm P)\times (S\pm P)$ operators, 
which correspond to the case of $G_3 = 0$.
The orange line is the plot for $(V- A)\times (V+ A)$ operator, which corresponds to the case of $G_0=0$.
The green and red lines represent the plots of $S\times S$ and $P\times P$ operators respectively.

\section{Interpretation of the magnetic dependence in each operator}
\label{sec3}

This section provides 
a possible explanation for the magnetic field dependence 
of the transition probability depicted in Fig.\ref{fig1}.
The qualitative behavior of the dependence in each operator can be understood by examining the spins
of muons and electrons involved in the transition process.
In the previous section, we presented the transition amplitudes as formulae derived from Ref.\cite{Fukuyama:2021iyw},
which are expressed using energy eigenstates in magnetic fields.
To qualitatively understand the magnetic field dependence in each operator,
we can analyze the transition amplitudes in the spin eigenstates of Mu and $\overline{\rm Mu}$
in relation to the energy eigenstates in the magnetic field.
Appendix \ref{appendix:A} presents a description of
the energy eigenstates in the magnetic field. 
Detailed algebraic calculations for the amplitudes in spin eigenstates can be found in Appendix \ref{appendix:C}.
In this section,
we extract essential points about magnetic field dependence in each operator,
focusing on spin conservation in the transition via pseudo-scalar and scalar exchanges.

\begin{figure}
\center
\includegraphics[width=6cm]{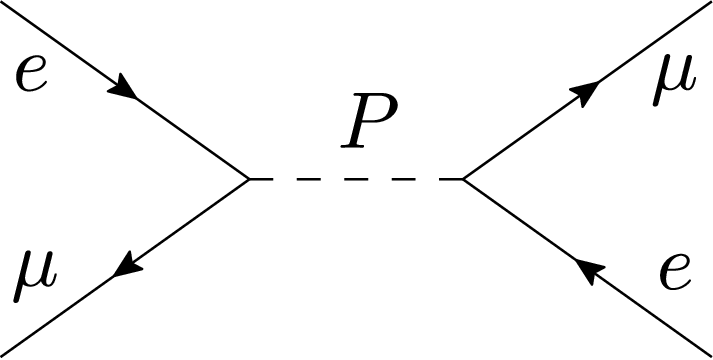}
\hspace{10mm}
\includegraphics[width=6cm]{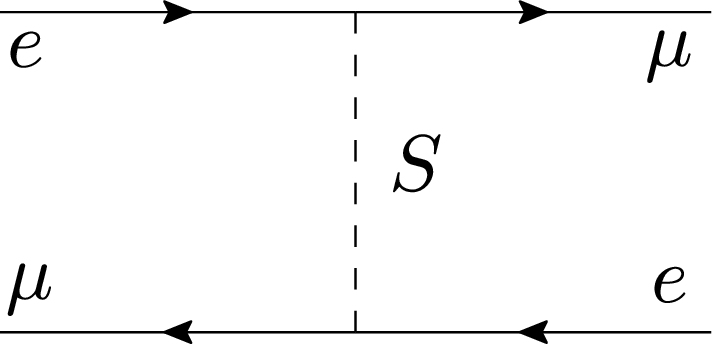}
\caption{$s$-channel exchange of a pseudo-scalar (left), and $t$-channel exchange of a scalar (right)
for the Mu-to-$\overline{\rm Mu}$ transition.}
\label{fig2}
\end{figure}

We first consider the processes via neutral (pseudo-)scalar exchange  
as depicted in Fig.\ref{fig2}.
The $s$-channel exchange of a pseudo-scalar (depicted on the left in Fig.\ref{fig2})
can generate the $P\times P$ operator in Eq.(\ref{eq:QP}) for the Mu-to-$\overline{\rm Mu}$ transition.
It should be noted that the $F=0$ singlet state of Mu is a pseudo-scalar.
The singlet state can transition to $\overline{\rm Mu}$ via the $s$-channel exchange of a pseudo-scalar.
On the other hand, the $F=1$ triplet state is a 3-vector that has even parity,
and the $s$-channel exchange of a pseudo-scalar cannot generate the transition for 
the $F=1$ triplet state.
Alternatively, we may focus on the spins of the muon and electron in Mu shown in the diagram.
Since their spins must be oriented oppositely, the process depicted in Fig.\ref{fig2} (left)
cannot involve the transition of the triplet state.
From Eq.(\ref{eq:QP}),
one can deduce that $G_0 +G_3/2 = 0$ for the $P\times P$ operator,
and ${\cal M}_{1,m} = 0$ can be confirmed in Eq.(\ref{eq:M1m}).
In Fig.\ref{fig1},
it can be observed that the $P \times P$ plot remains flat until the $m=0$ states experience the magnetic field
at $B \sim 0.01\,$T.
This absence of the triplet transition amplitude prevents the triplet state from transitioning, 
even in a weak magnetic field.

The transition via the $S\times S$ operator
corresponds to the $t$-channel exchange of a scalar (on the right in Fig.\ref{fig2}).
{}From the diagram, one finds that
the spins of $\mu^+$ in Mu and $e^+$ in $\overline{\rm Mu}$
are the same,
as 
are the spins of $e^-$ in Mu and $\mu^-$ in $\overline{\rm Mu}$.
This implies that $|{\rm Mu}; \uparrow \downarrow \rangle$ transitions 
to $|\overline{\rm Mu}; \downarrow \uparrow \rangle$,
and $|{\rm Mu}; \downarrow \uparrow \rangle$ transitions
to $|\overline{\rm Mu}; \uparrow \downarrow \rangle$,
where
we denote the spins of Mu and $\overline{\rm Mu}$ with up and down arrows in the order of muons and electrons.
As derived in Appendix \ref{appendix:A},
the $m=0$ energy eigenstates in a magnetic field are given by
\begin{align}
\left(
\begin{array}{c} 
 |{\rm Mu}; 1,0 \rangle_B \\
|{\rm Mu};0,0 \rangle_B 
\end{array}  
\right)
&
=
 \left(
\begin{array}{cc}
c & s \\
-s & c 
\end{array}
\right)
\left(
\begin{array}{c} 
|{\rm Mu}; \downarrow \uparrow \rangle \\
|{\rm Mu}; \uparrow \downarrow \rangle 
\end{array}  
\right) , 
\label{eq:MuB}
\\
\left(
\begin{array}{c} 
 |\overline{\rm Mu}; 1,0 \rangle_B \\
|\overline{\rm Mu};0,0 \rangle_B 
\end{array}  
\right)
&
=
 \left(
\begin{array}{cc}
s & c \\
-c & s 
\end{array}
\right)
\left(
\begin{array}{c} 
|\overline{\rm Mu}; \downarrow \uparrow \rangle \\
|\overline{\rm Mu}; \uparrow \downarrow \rangle 
\end{array}  
\right) ,
\label{eq:MubarB}
\end{align}
where
\begin{equation}
c  = \frac1{\sqrt2} \sqrt{ 1+ \frac{X}{\sqrt{1+X^2}} },
\qquad
s  = \frac1{\sqrt2} \sqrt{ 1- \frac{X}{\sqrt{1+X^2}} }.
\end{equation}
One finds that
\begin{align}
{\cal M}_{1,0}^B &= {}_B\langle \overline{\rm Mu} ; 1,0 | Q_S |
{\rm Mu}; 1,0 \rangle_B
=
s^2 \langle \overline{\rm Mu} ; \downarrow \uparrow |Q_S |
{\rm Mu} ;  \uparrow \downarrow \rangle 
+ c^2
\langle \overline{\rm Mu} ; \uparrow \downarrow |Q_S |
{\rm Mu} ;  \downarrow \uparrow \rangle, \\
{\cal M}_{0,0}^B &= {}_B\langle \overline{\rm Mu} ; 0,0 | Q_S |
{\rm Mu}; 0,0 \rangle_B
=
-c^2 \langle \overline{\rm Mu} ; \downarrow \uparrow |Q_S |
{\rm Mu} ;  \uparrow \downarrow \rangle 
- s^2
\langle \overline{\rm Mu} ; \uparrow \downarrow |Q_S |
{\rm Mu} ;  \downarrow \uparrow \rangle 
\end{align}
are derived from spin conservation in the process.
Note 
that the consistency of the amplitude for $B=0$
requires
$\langle \overline{\rm Mu} ; \downarrow \uparrow |Q_S |
{\rm Mu} ;  \uparrow \downarrow \rangle 
=
\langle \overline{\rm Mu} ; \uparrow \downarrow |Q_S |
{\rm Mu} ;  \downarrow \uparrow \rangle$,
which can actually be obtained in Eq.(\ref{eq:app-Qs-spin}) through properly configured calculations.
Consequently, it can be deduced that
 the transition amplitudes do not depend on the magnetic field due to $c^2+s^2 = 1$.
Furthermore, ${\cal M}_{1,0}^B = - {\cal M}_{0,0}^B$.
In fact, Eq.(\ref{eq:QS}) demonstrates $G_0- G_3/2 =0$ for the $S\times S$ operator,
and one can verify that the amplitudes do not depend on the magnetic field, along with ${\cal M}_{1,0}^B = - {\cal M}_{0,0}^B$
from Eqs.(\ref{eq:M10B}) and (\ref{eq:M00B}).
As a result, the $S\times S$ plot remains flat in the strong magnetic field, 
as can be seen in Fig.\ref{fig1}.

\begin{figure}
\center
\includegraphics[width=6cm]{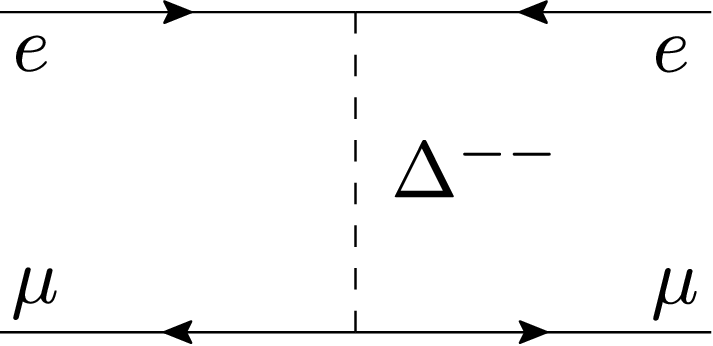}
\hspace{10mm}
\includegraphics[width=6cm]{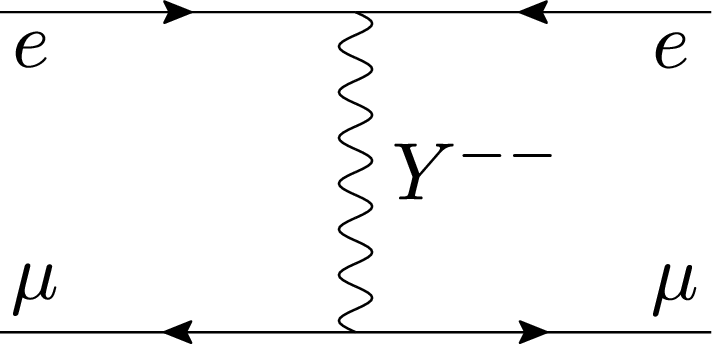}
\caption{$t$-channel exchange of a doubly charged scalar (left), and vector boson (right).}
\label{fig3}
\end{figure}

Next, let us consider the Mu-to-$\overline{\rm Mu}$ transition via the $t$-channel exchange of doubly charged particles
as shown in Fig.\ref{fig3}.
Through the doubly charged scalar exchange \cite{Chang:1989uk,Swartz:1989qz},
a transition operator $(\bar \mu \mu^c)(\bar {e^c} e)$ is induced.
The operator can include $Q_1,Q_2,Q_4$, and $Q_5$ (though the $Q_4, Q_5$ contributions through the exchange are expected 
to be small
in the $SU(2)_L \times U(1)_Y$ theory).
In the case of the scalar exchange, the spins of $\mu^+$ in Mu and $\mu^-$ in $\overline{\rm Mu}$ are the same,
as are the spins of $e^-$ in Mu and $e^+$ in $\overline{\rm Mu}$,
as found from the Feynman diagram in Fig.\ref{fig3}.
By considering spin conservation and Eqs.(\ref{eq:MuB}) and (\ref{eq:MubarB}), one can obtain
\begin{align}
{\cal M}_{1,0}^B &= {}_B\langle \overline{\rm Mu} ; 1,0 | Q_1 |
{\rm Mu}; 1,0 \rangle_B
=
sc \left( \langle \overline{\rm Mu} ; \downarrow \uparrow |Q_1 |
{\rm Mu} ;  \downarrow \uparrow \rangle 
+ 
\langle \overline{\rm Mu} ; \uparrow \downarrow |Q_1 |
{\rm Mu} ;  \uparrow \downarrow \rangle \right), \\
{\cal M}_{0,0}^B &=
{}_B\langle \overline{\rm Mu} ; 0,0 | Q_1 |
{\rm Mu}; 0,0 \rangle_B
=
sc \left( \langle \overline{\rm Mu} ; \downarrow \uparrow |Q_1 |
{\rm Mu} ;  \downarrow \uparrow \rangle 
+
\langle \overline{\rm Mu} ; \uparrow \downarrow |Q_1 |
{\rm Mu} ;  \uparrow \downarrow \rangle \right) ,
\end{align}
and ${\cal M}_{1,0}^B = {\cal M}_{0,0}^B \propto 1/\sqrt{1+X^2}$.
As pointed out in Appendix \ref{appendix:A}, 
the spin orientations need to be reversed for the Mu-to-$\overline{\rm Mu}$ transition in a strong magnetic field,
and due to the spin conservation, 
the transition via doubly charged scalar exchange will not take place
in the limit of a strong magnetic field.
Indeed, this can be verified in 
the transition probabilities for $G_3 = 0$ in Eqs.(\ref{time-dependent-B2}), 
and the blue line for the $(V\pm A) \times (V\pm A)$ and $(S\pm P) \times (S\pm P)$ operators in Fig.\ref{fig1}
touches zero for $B \agt 1\,$T.
In the case of doubly charged vector boson exchange \cite{Frampton:1989fu,Fujii:1992np,Frampton:1997in},
a transition operator $(\bar \mu \gamma_\alpha \mu^c)(\bar {e^c} \gamma^\alpha e)$ 
is generated. 
The induced operator is proportional to the $(V-A) \times (V+A)$ operator $Q_3$, 
which can be obtained from Fierz transformation.
In this case, the spins of the muons and electrons in Mu and $\overline{\rm Mu}$ can either be the same or opposite, 
and the transition can occur in the limit of a strong magnetic field,
as can be seen from the the orange line in Fig.\ref{fig1}.

We have found that the amplitude of the triplet state vanishes for the $P\times P$ operator initially.
Then, one may wonder when the transition amplitude of the singlet state becomes zero, 
that is, when ${\cal M}_{0,0} = 0$.
Considering that the amplitude of the triplet state disappears due to $s$-channel pseudo-scalar exchange,
 a case involving $s$-channel vector exchange is a candidate.
However, due to the Lorentz invariance, $s$-channel neutral vector boson exchange introduces an additional term.
For example, one can contemplate combining $S \times S$ with $V\times V$ operator to cancel the additional term.
Another possibility involves considering $2Q_S + Q_P$.
The rationale behind the vanishing of the singlet amplitude in these combinations 
can be comprehended by referring to
Eqs.(\ref{eq:MQS}), (\ref{eq:MQV}) and (\ref{eq:3-vector-formula}).
In any case, it appears that the disappearance of the singlet amplitude cannot be attributed to a single Lorentz invariant operator.

\section{What can the measurements of probability ratios tell us?}
\label{sec4}

In this section, we define ratios of the transition probabilities 
under different magnetic fields,
and we elucidate the insights that can be inferred from these ratios,
which would be measured at the MACE and J-PARC experiments.

Employing Eqs.({\ref{eq:f10-f00}) for $f_{1,0}$ and $f_{0,0}$,
we obtain the ratio of the transition probabilities from Eqs.(\ref{time-dependent-B1}) and (\ref{time-dependent-B2})
for the magnetic fields $B\agt O(10)\, \mu$T and $B=B_0 \ll 1\, \mu$T,
resulting in
\begin{equation}
 \frac{P(t, B)}{P(t, B_0)}
= \frac{4 |G_0|^2 - 4\,{\rm Re} (G_0 G_3^*) (1+2P_\mu X)+ |G_3|^2 (5+4P_\mu X + 4X^2)}{8(1+X^2)(|G_0|^2+ \frac34 |G_3|^2)},
\end{equation}
where $P_\mu$ is the muon polarization in the produced Mu.

In the context of the J-PARC experiment, if the transitions are observed, 
we anticipate measuring the probability ratio between
$B= B_0 \ll 1\, \mu{\rm T}$ and $B= B_1$ (where $O(10)\, \mu{\rm T} < B_1 < O(1)\,{\rm mT}$, i.e., $X\ll 1$),
denoted as $R_1$,
\begin{equation}
R_1 \equiv  \frac{P(t,B_1)}{P(t,B_0)} = 
\frac{|M_{1,m}|^2 + |M_{0,0}|^2}{3|M_{1,m}|^2 + |M_{0,0}|^2} = \frac12 \frac{|G_0|^2 - {{\rm Re}\,G_0 G_3^*}+ \frac54 |G_3|^2}{|G_0|^2 + \frac34 |G_3|^2}. 
\label{PtB-ratio}
\end{equation}
Collecting data from both the MACE and J-PARC experiments, 
we anticipate obtaining the ratio $R_2$ under a stronger magnetic field $B_2$,
\begin{equation}
R_2 \equiv \frac{P(t,B_2)}{P(t,B_0)}.
\end{equation}
In the MACE experiment, the magnetic field strength will be $B_2 = 0.1\,$T (corresponding to $X = 0.63$).
Given the distinct methods used in these two experiments,
it is crucial to convert the time-integrated probability measured by MACE
into a time-dependent probability format within the framework of the J-PARC methods.
This conversion is necessary to calculate the ratio $R_2$.

As we have explained, 
doubly charged mediators can produce either $G_0$ or $G_3$,
while neutral mediators can generate both $G_0$ and $G_3$ elements at the tree level.
In general,
the complex phases of $G_0$ and $G_3$ can be different.
It is crucial to note
that 
the relative phase between $G_0$ and $G_3$ must be extremely small due to the electron EDM
if a single neutral mediator is responsible for generating the transition operator.
This will be explored further in the subsequent sections, where we delve 
into concrete models to provide a clearer understanding.

Suppose that the relative phase is zero (i.e., ${\rm Im}\, G_3/G_0 =0)$. 
The measurement of the ratio $R_1$ at J-PARC
can provide two potential solutions for $G_3/G_0$
(except for the maximum (minimum) value of $R_1 = 1$ ($=1/3$))
from Eq.(\ref{PtB-ratio}).
If the ratio $R_2$ is also obtained,
the true solution can be distinguished, and both $G_3$ and $G_0$ can be determined in principle.
Additionally, 
the value of the muon polarization $P_\mu$ in the produced Mu can be determined (unless $R_1 = 1/2$), 
as illustrated in Fig.\ref{fig4} (top).
The intersections of lines on the figure for different $P_\mu$ values
correspond to points for the $S\times S$ operator ($R_1 = R_2 = 1/2$) 
and the case of $G_3 = 0$ ($R_1 = 1/2$, $R_2 = 1/2(1+X^2)$).
The values of $R_2$ for these two points are independent of $P_\mu$ due to
  $|{\cal M}_{1,0}^B|=  |{\cal M}_{0,0}^B|$ and $f_{1,0} + f_{0,0} = 1/2$.
If the muon polarization $P_\mu$ in Mu at $B=B_2$ can be accurately measured experimentally,
it becomes possible to investigate whether $G_3/G_0$ has a phase using the values of $R_1$ and $R_2$,
as can be understood from Fig.\ref{fig4} (bottom).

We comment on possible further analyses 
if the transition is really observed at the experiments.
In the J-PARC method, 
a narrower laser band can be 
used 
to ionize only the $F=1$ states of $\overline{\rm Mu}$
at a weak magnetic field.
If such selective ionization is possible, 
the J-PARC experiment by itself can determine $G_0$ and $G_3$
as well as their relative physical phase
without information of the muon polarization in the produced Mu.
If measurements at stronger magnetic fields
can be performed in the PSI method, 
it provides a cross-check for the determination of $G_0$ and $G_3$.

\begin{figure}
\center
\begin{tabular}{c}
\includegraphics[width=10cm]{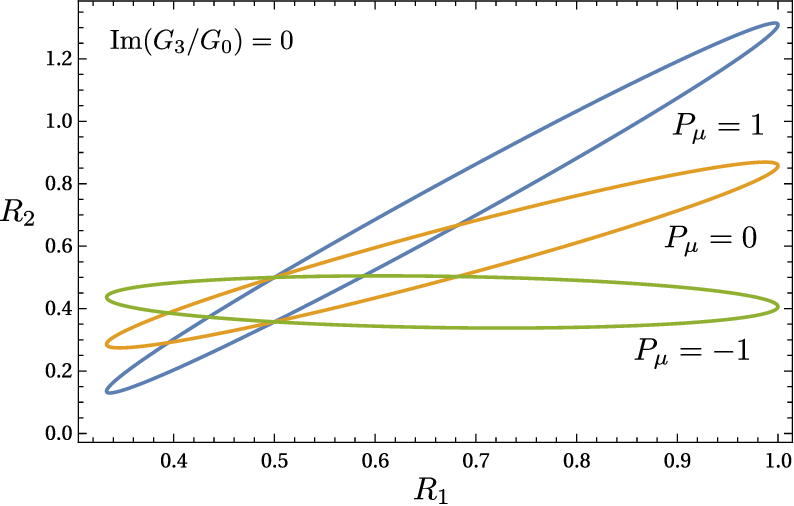}
\\
\\
\\
\includegraphics[width=10cm]{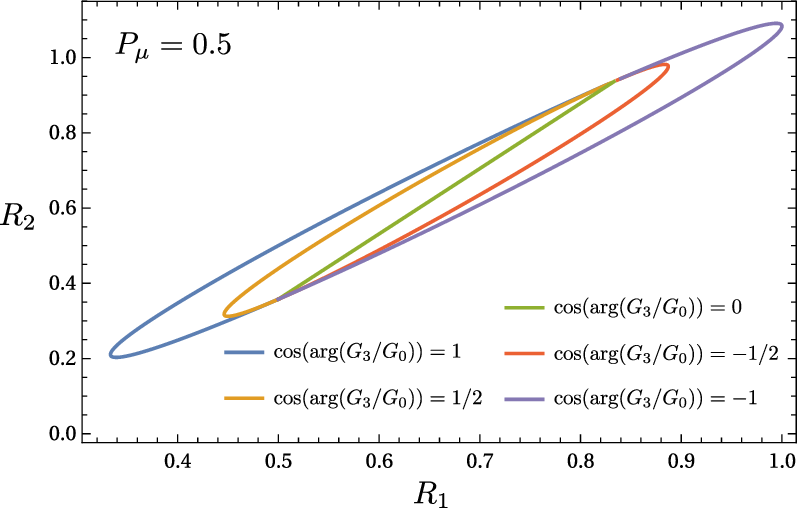}
\end{tabular}
\vspace{1cm}
\caption{
Trajectories in the $R_1$-$R_2$ plain while varying the value of $G_3/G_0$
under ${\rm Im} \,G_3/G_0 = 0$ with
different values of muon polarization $P_\mu$ in the produced Mu (top),
and 
trajectories for different phases of $G_3/G_0$
with a fixed muon polarization $P_\mu = 0.5$ (bottom).
}
\label{fig4}
\end{figure}

\section{Axion-like particle}
\label{sec5}

In this section, we examine a model incorporating an
axion-like particle (ALP).
While the intricate specifics of the ALP are discussed
in Refs.\cite{Bauer:2019gfk,Endo:2020mev,Calibbi:2020jvd},
we focus on the following effective Lagrangian presented in Ref.\cite{Endo:2020mev},
which deals with the Mu-to-$\overline{\rm Mu}$ transition via the ALP:
\begin{equation}
{\cal L} = (y_V\, a \,\bar \mu e + y_A \, a \,\bar \mu \gamma_5 e + {\rm H.c.}) - \frac12 m_a^2 a^2.
\end{equation}
Here, these $y_V$ and $y_A$ couplings originate from vector and axial-vector couplings with the ALP $a$, respectively.
We have made the assumption that 
 the couplings of the ALP do not give rise to $\Delta L_e = - \Delta L_\mu = \pm 1$ processes. 
By integrating out the ALP,
we obtain
\begin{equation}
{\cal L} \supset \frac{1}{2 m_a^2} \left( y_V^2 (\bar \mu e)^2 + y_A^2 (\bar \mu \gamma_5  e)^2+ 2 y_V y_A (\bar \mu e)(\bar \mu \gamma_5 e) \right).
\end{equation}
It is worth noting that the final term is proportional to $Q_4 - Q_5$ and
has no impact on $G_0$.
We obtain
\begin{equation}
G_0 = \frac{1}{8\sqrt2} \frac{y_V^2 + y_A^2}{m_a^2},
\qquad
G_3 = \frac{1}{4\sqrt2} \frac{y_V^2 - y_A^2}{m_a^2}.
\label{eq:G0G3-ALP}
\end{equation}

The experimental bound of the Mu-to-$\overline{\rm Mu}$ transition 
places constraints on the magnitude of $y_V^2/m_a^2$ for a given $y_V/y_A$,
and consequently, the contribution to muon $g-2$ cannot be sufficiently significant \cite{Endo:2020mev}
to account for the deviation between the theoretical prediction in SM and the experimental measurement \cite{Aoyama:2020ynm,Muong-2:2023cdq}.
If there is no flavor violation, one finds a relationship between the electron and muon $g-2$
for their contributions from new physics,
and thus, the contribution to the electron $g-2$ is also small.
However, due to the violation of lepton flavor by the ALP couplings,
the electron $g-2$ has an additional contribution \cite{Nie:1998dg,Galon:2016bka}.
By ignoring the term of $O(m_e^2/m_a^2)$, we obtain the contribution to the electron $g-2$ 
(using a widely adopted convention, 
we denote $\Delta a_e$ as the new physics contribution to $a_e \equiv (g_e-2)/2$, with $g_e$ representing the $g$-factor of the electron) as
\begin{equation}
\Delta a_e =  \frac{1}{16\pi^2} \frac{m_e m_\mu}{m_a^2}
(|y_V|^2 - |y_A|^2)\,  f\!\left(\frac{m_a^2}{m_\mu^2} \right) ,
\label{eq:Dae-ALP}
\end{equation}
where $f(x)$ is a loop function, $f(x) = (2x^3 \ln x - 3x^3 + 4x^2 -x)/(x-1)^3$,
which is positive for any $x\,(>1)$.

The electron EDM is given as
\begin{equation}
d_e = \frac{e m_\mu}{32\pi^2 m_a^2} {\rm Im}(y_V y_A^*) f\!\left(\frac{m_a^2}{m_\mu^2} \right).
\label{eq:EDM-ALP}
\end{equation}
One finds that ${\rm Im}(y_V y_A^*)$ must be extremely small to satisfy the experimental bound of the electron EDM \cite{ACME:2018yjb,Roussy:2022cmp}.
If $y_V/y_A$ is real, both $y_V$ and $y_A$ can be made to be real without loss of generality
 by unphysical phase rotation
of $\mu$ and $e$ fields.
We will consider $y_V$ and $y_A$ as real values from this point onward.
Then, there is no physical phase present in the transition amplitudes,
as obviously found from Eq.(\ref{eq:G0G3-ALP}).

The ratio of the transition probability in Eq.(\ref{PtB-ratio}) is 
\begin{equation}
R_1 = \frac{y_V^4 - 2y_V^2 y_A^2 + 2 y_A^4}{2(y_V^4 - y_V^2 y_A^2 + y_A^4)}
= \frac12 + \frac{y_A^2(y_A^2-y_V^2)}{2(y_V^4 - y_V^2 y_A^2 + y_A^4)}.
\label{eq:R1-ALP}
\end{equation}
One can confirm that $R_1 = 1/2$ if $y_A = 0$, and $R_1 = 1$ if $y_V = 0$,
reflecting the $S\times S$ and $P\times P$ cases, respectively.
Upon closer examination of this equation, it becomes apparent that
$R_1 > 1/2$ if $y_A^2  > y_V^2$, and $R_1 \leq 1/2$ if $y_V^2 > y_A^2$.
We immediately find from Eq.(\ref{eq:Dae-ALP}) that
a negative $\Delta a_e$ leads to $R_1 > 1/2$,
while
a positive $\Delta a_e$ results in
$R_1 \leq 1/2$.
This prediction, significant within the scope of this model,
can be tested through the measurements of the Mu-to-$\overline{\rm Mu}$ transition
at J-PARC.

\begin{figure}[p]
\center
\includegraphics[width=8cm]{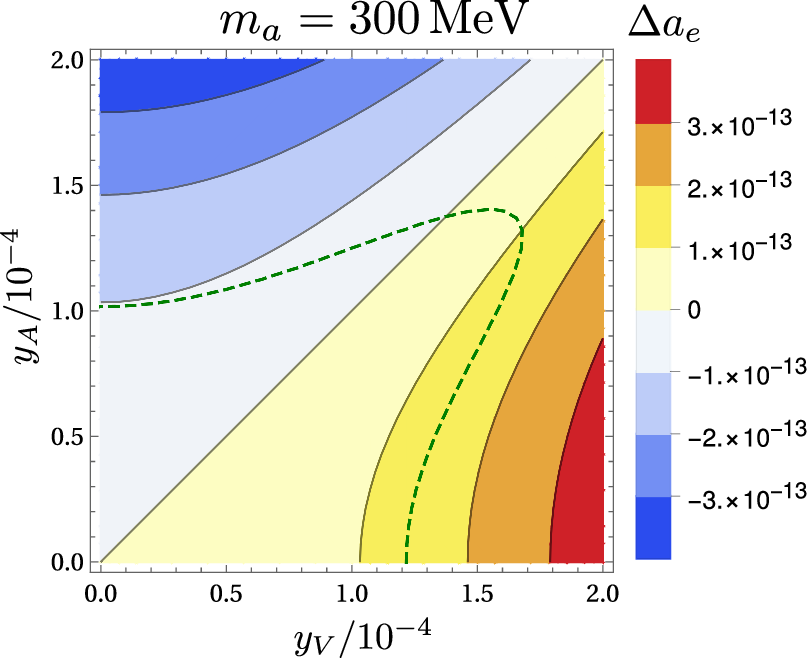}
\hspace{5mm}
\includegraphics[width=8cm]{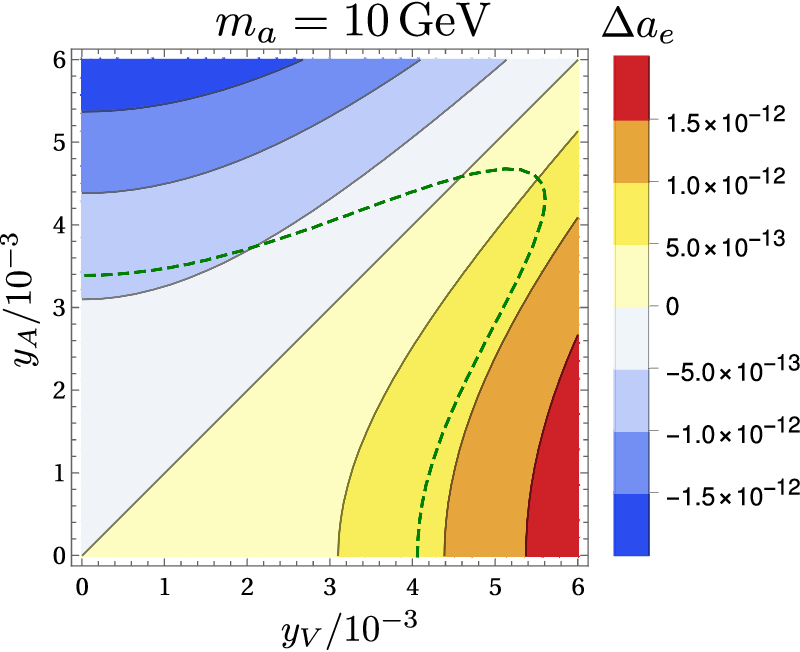}
\caption{
Contour plots illustrating $\Delta a_e$ for two ALP masses, $m_a = 300$ MeV and $m_a = 10$ GeV.
The green dashed lines delineate contours of
 $\bar P = 8.3 \times 10^{-11}$, which correspond to the bound of the Mu-to-$\overline{\rm Mu}$ transition
 given by the PSI experiment. 
}
\label{fig5}
\end{figure}

\begin{figure}[p]
\center
\includegraphics[width=10cm]{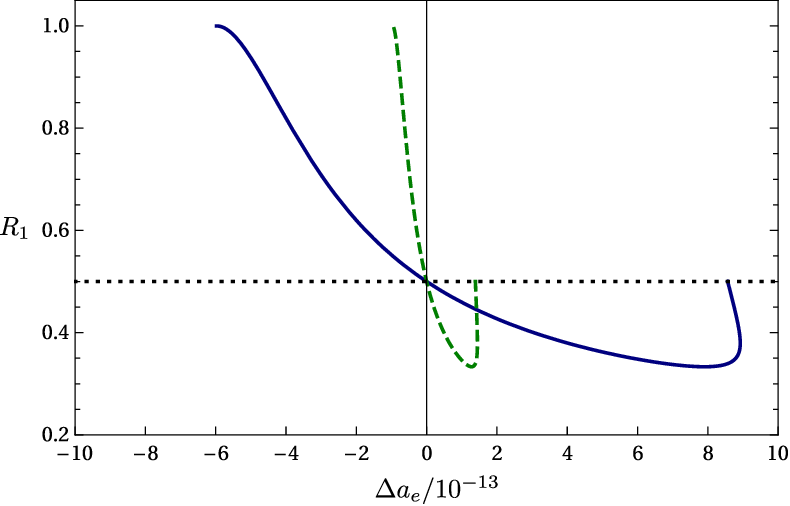}
\caption{
Plot depicting the relation between the contribution to the electron $g-2$ ($\Delta a_e$) 
and the transition probability ratio $(R_1)$ in the ALP model for $m_a = 300\,$MeV 
(indicated by the green dashed line) and
$m_a = 10\,$GeV (indicated by the blue solid line) when the time-integrated transition probability
is just same as the PSI bound, $\bar P = 8.3 \times 10^{-11}$. 
The horizontal dotted line corresponds to $R_1 = 1/2$.
}
\label{fig5-2}
\end{figure}

We provide an overview of the current status concerning the electron $g-2$.
In 2018, the measurement of the fine structure constant employing Cs exhibited enhanced precision \cite{Parker:2018vye},
thus highlighting a discrepancy between the experimental measurements and the
theoretical calculations in SM, even in the electron $g-2$.
In 2020, the fine structure constant was measured using Rb atoms \cite{Morel:2020dww}.
However, these two measurements of the fine structure constant displayed a $5\sigma$ discrepancy.
Recently, the experimental measurement of the electron $g-2$ was updated \cite{Fan:2022eto}.
The numerical values of the deviation of $a_e =(g_e-2)/2$ between the experimental measurement
and theoretical calculations utilizing the measurements of the fine structure constant 
are as follows:
\begin{align}
\Delta a_e^{\rm Cs} &= a_e ({\rm exp.}; 2022) - a_e ({\rm SM, Cs}; 2018) = (-1.02 \pm 0.26) \times 10^{-12}, \label{eq:e-gminus2-Cs}\\
\Delta a_e^{\rm Rb} &= a_e ({\rm exp.}; 2022) - a_e ({\rm SM, Rb}; 2020) = (0.34 \pm 0.16) \times 10^{-12}.\label{eq:e-gminus2-Rb}
\end{align}
The $5\sigma$ discrepancy between the Cs and Rb measurements of the fine structure constant 
is crucial to address the potential presence 
of new physics contributions.
At present, the sign of $\Delta a_e$ remains indeterminate.

The bound of the Mu-to-$\overline{\rm Mu}$ transition
places a restriction on $(y_V^2-y_A^2)/m_a^2$ for a fixed $y_V/y_A$ value. 
Given the presence of a logarithmic factor in the loop function:
\begin{equation}
f \left( \frac{m_a^2}{m_\mu^2} \right)
= 2 \ln \frac{m_a^2}{m_\mu^2} - 3 \ \ ( m_a \gg m_\mu ) ,
\label{eq:log-enhance-scalar}
\end{equation}
it becomes apparent that $\Delta a_e$ can attain greater magnitudes with increasing ALP mass.

In Fig.\ref{fig5}, we present the contours of $\Delta a_e$
as functions of the ALP couplings $y_V$ and $y_A$.
The green dashed line signifies the current bound resulting from the PSI transition experiment.
For the purpose of plotting the dashed lines,
we choose $P_\mu = 0.5$ as the muon polarization in the produced Mu.
In the case of $m_a = 300\,$MeV, $|\Delta a_e|$ remains modest 
and does not fall within the $1\sigma$ range given in
Eqs.(\ref{eq:e-gminus2-Cs}) and (\ref{eq:e-gminus2-Cs}).
In the case of $m_a = 10\,$MeV, on the other hand,
$|\Delta a_e|$ can become larger as anticipated earlier.
Notably, the viability of the ALP  for $m_a < 10\,$GeV can be assessed through the Belle II experiment \cite{Endo:2020mev}.

In Fig.\ref{fig5-2},
we present the relation between $\Delta a_e$ and the ratio $R_1$ when the transition probability
is just same as the PSI bound, $\bar P = 8.3 \times 10^{-11}$.
The blue solid line corresponds to the case of $m_a=10\,$GeV, while
the green dashed line represents the case of $m_a = 300\,$MeV.
As explained earlier,
the heavier ALP allows a larger magnitude of $|\Delta a_e|$. 
The value of $|\Delta a_e|$ with fixed values of $m_a$ and $y_V/y_A$ is proportional to $\sqrt{\bar P}$.
Consequently, 
once the MACE achieves its targeted goal of $\bar P \sim O(10^{-14})$ \cite{Bai:2022sxq},
it will be become feasible to determine whether this model contributions to the electron $g-2$. 
As explained earlier from (\ref{eq:Dae-ALP}) and (\ref{eq:R1-ALP}),
this figure underscores 
that $R_1 \leq 1/2$ when $\Delta a_e$ is positive
and $R_1 > 1/2$ when $\Delta a_e$ is negative.
These findings will be pivotal if the transition is actually observed
and the probability ratio is measured at J-PARC.

\section{Inert doublet model}
\label{sec6}

The Mu-to-$\overline{\rm Mu}$ transition 
can potentially arise from the inclusion of an additional $SU(2)_L$ doublet \cite{Hou:1995dg}.
This type of the model can be also contemplated 
within the framework of $R$-parity violating supersymmetry \cite{Halprin:1993zv}.
In this section, we consider an inert $SU(2)_L$ doublet, one that remains devoid of a vacuum expectation value,
and thus preserves a symmetry aimed at diminishing
$\Delta L_e = -\Delta L_\mu = \pm 1$ processes.
Similar models that give rise to the Mu-to-$\overline{\rm Mu}$ transition are also considered in recent works \cite{Afik:2023vyl,Heeck:2023iqc}.

We consider the SM Higgs doublet $\Phi$, 
accompanied by an inert doublet $\eta$ which does not acquire a vacuum expectation value:
\begin{equation}
\Phi = \left(
\begin{array}{c}
 \omega^+
\\
 v + \frac{h+i \omega^0}{\sqrt2} 
 \end{array}
\right),
\qquad
\eta = \left(
\begin{array}{c}
 H^+
\\
 \frac{H+i A}{\sqrt2} 
 \end{array}
\right),
\end{equation}
where $\omega^+$ and $\omega^0$ represent 
Nambu-Goldstone bosons that would be absorbed by the $W$ and $Z$ bosons. 
The physical Higgs boson with a mass of 125 GeV corresponds to $h$. 
Our model revolves around the utilization of a global discrete $Z_4$ symmetry,
where the following charges are assigned to the left-handed lepton doublets, right-handed charged leptons,
and scalar doublets $\Phi$ and $\eta$: 
\begin{equation}
\ell_e : 1,
\quad
\ell_\mu : 3,
\quad
\ell_\tau : 0,
\quad
e_R : 1,
\quad
\mu_R : 3,
\quad
\tau_R : 0,
\quad
\Phi : 0,
\quad
\eta : 2,
\label{eq:discrete-symmetry}
\end{equation}
leading to permissible LFV couplings as 
\begin{equation}
-{\cal L} =   \rho_{12}\, \overline{\ell_e}\, \eta\,   \mu_R +  \rho_{21}\, \overline{\ell_\mu}\, \eta\,   e_R + {\rm H.c.}.
\label{eq:Yukawa-eta}
\end{equation}
The scalar potential terms are
\begin{equation}
V = m_\Phi^2 \Phi^\dagger \Phi + m_\eta^2 \eta^\dagger \eta + \lambda_1 (\Phi^\dagger \Phi)^2 
+ \lambda_2 (\eta^\dagger \eta)^2
+ \lambda_3 \Phi^\dagger \Phi \eta^\dagger \eta
+ \lambda_4 \eta^\dagger \Phi \Phi^\dagger \eta
+\frac{\lambda_5}{2} \left(  (\Phi^\dagger \eta)^2 + {\rm H.c.}  \right),
\end{equation}
and the masses of the scalars in $\eta$
are
\begin{align}
m_H^2 &= m_\eta^2 + (\lambda_3  + \lambda_4  + \lambda_5) v^2, \\
m_A^2 &= m_\eta^2 + (\lambda_3  + \lambda_4  - \lambda_5) v^2, \\
m_{H^+}^2 &= m_\eta^2 + \lambda_3   v^2.
\end{align}
This discrete symmetry arrangement engenders 
the absence of any mixing between $h$ and $H$;
in other words, $h$ and $H$ are brought into alignment (for instance, the couplings of gauge bosons and $h$ remain consistent with those of SM) without decoupling $H$.
We note that this discrete symmetry is not spontaneously broken and remains intact 
even after the electroweak symmetry breaking,
and the lightest scalar in the $\eta$ doublet decays into two leptons via the interaction in Eq.(\ref{eq:Yukawa-eta}).

The discrete charge assignments in Eq.(\ref{eq:discrete-symmetry})
forbids the $\Delta L_e = -\Delta L_\mu = \pm 1$ processes,
while corrections to the muon and electron masses are possible,
and the muon and electron $g-2$ can be generated as
\begin{align}
\Delta a_\mu &\simeq
\frac{m_\mu^2}{16\pi^2} \frac16 \left[\left(|\rho_{12}|^2+|\rho_{21}|^2 \right)\left(\frac{1}{m_{H}^2} + \frac{1}{m_A^2} \right) - |\rho_{12}|^2 \frac{1}{m_{H^+}^2}\right], \\
\Delta a_e &\simeq
\frac{m_e^2}{16\pi^2} \frac16 \left[\left(|\rho_{12}|^2+|\rho_{21}|^2 \right)\left(\frac{1}{m_{H}^2} + \frac{1}{m_A^2} \right) - |\rho_{21}|^2 \frac{1}{m_{H^+}^2}\right] \nonumber \\ 
&+ \frac{m_e m_\mu}{16\pi^2}\, {\rm Re} (\rho_{12}\rho_{21})\left(\left( \ln \frac{m_H^2}{m_\mu^2}-\frac32 \right) \frac{1}{m_H^2}
-
\left(  \ln \frac{m_A^2}{m_\mu^2}-\frac32 \right) \frac{1}{m_A^2}
\right).
\end{align}
The electron EDM is obtained as
\begin{align}
 d_e \simeq
 \frac{e m_\mu}{32\pi^2} \,{\rm Im} (\rho_{12}\rho_{21})\left(\left( \ln \frac{m_H^2}{m_\mu^2}-\frac32 \right) \frac{1}{m_H^2}
-
\left(  \ln \frac{m_A^2}{m_\mu^2}-\frac32 \right) \frac{1}{m_A^2}
\right).
\end{align}
The experimental constraints on the electron EDM 
restrict the model parameters such that they satisfy either ${\rm Im} (\rho_{12}\rho_{21}) \simeq 0$ or $m_H = m_A$.
Without loss of generality, one can make either $\rho_{12}$ or $\rho_{12}$ (as well as $\lambda_5$) real 
through unphysical phase rotations of fields while ensuring that the electron and muon masses remain real.
Considering the constraints on the electron EDM when $m_H \neq m_A$,
 we will suppose that both $\rho_{12}$ and $\rho_{21}$ are real numbers.

Noting 
\begin{align}
 \rho_{21} \overline{\mu_L} e_R H
+ \rho_{12} \overline{e_L} \mu_R H + {\rm H.c.} 
=
\rho_+ \bar\mu e H
+ \rho_- \bar\mu \gamma_5 e H
+ {\rm H.c.},  \\
i \rho_{21} \overline{\mu_L} e_R A
+ i\rho_{12} \overline{e_L} \mu_R A + {\rm H.c.} 
=
i \rho_- \bar\mu e A
+ i\rho_+ \bar\mu \gamma_5 e A
+ {\rm H.c.} ,
\end{align}
where
\begin{equation}
\rho_+ \equiv \frac{\rho_{21}+\rho_{12}^*}{2},
\qquad
\rho_- \equiv \frac{\rho_{21}-\rho_{12}^*}{2},
\end{equation}
we obtain the $S\times S$ and $P \times P$ transition operators
by integrating out $H$ and $A$ scalar fields:
\begin{equation}
{\cal L} \supset \frac14 \left[
\left( \frac{\rho_+^2}{m_H^2} - \frac{\rho_-^2}{m_A^2} \right) (\bar \mu e)^2 
+ \left( \frac{\rho_-^2}{m_H^2} - \frac{\rho_+^2}{m_A^2} \right) (\bar \mu \gamma_5 e)^2 
\right].
\end{equation}
It is also convenient to express as
\begin{align}
G_0 &= \frac{1}{32\sqrt2} (\rho_{21}^2 + \rho_{12}^{*2}) \left(\frac{1}{m_H^2} - \frac{1}{m_A^2} \right), \\
G_3 &= \frac{1}{8\sqrt2} \rho_{21} \rho_{12}^* \left(\frac{1}{m_H^2} + \frac{1}{m_A^2} \right).
\end{align}
As previously explained, 
assuming ${\rm Im} (\rho_{12} \rho_{21}) = 0$ due to the electron EDM constraint
leads to the absence of a physical phase in the transition amplitudes.
Alternatively, 
the electron EDM can also be eliminated by setting $m_H = m_A$,
which also results in no physical phase in the amplitude because of $G_0=0$.

In the preceding ALP model,
the contribution to the muon $g-2$ is suppressed
due to the constraints imposed by the Mu-to-$\overline{\rm Mu}$ transition.
In the inert doublet model,
the transition amplitudes can be canceled 
by choosing
$m_H \simeq m_A$ (which means $\lambda_5 \to 0$) and either $|\rho_{12}| \ll |\rho_{21}|$ or $|\rho_{12}| \gg |\rho_{21}|$.
This choice of the parameters yields the contribution to the muon $g-2$ to be $\Delta a_\mu \sim 10^{-9}$ for $\rho_{21} \sim O(1)$ (or $\rho_{12} \sim O(1)$) \cite{Afik:2023vyl,Heeck:2023iqc}.
In this case, however, the contribution to the electron $g-2$ becomes diminished due to $m_H \simeq m_A$.

The contribution to the electron $g-2$ can become substantial with a value of $\lambda_5 \sim 1$, 
while this choice results in a small impact on the muon $g-2$.
Remarking
\begin{equation}
{\rm Re}(\rho_{12} \rho_{21})
= 
\left| \rho_+
\right|^2
-
\left| \rho_-
\right|^2,
\end{equation}
one can derive algebraically that 
$\Delta a_e$ is positive for
an $S\times S\,$-like transition operator
(with $\rho_-^2 m_A^2 \simeq \rho_+^2 m_H^2$),
and
$\Delta a_e$ is negative 
for a $P\times P\,$-like transition operator
(with $\rho_+^2 m_A^2 \simeq \rho_-^2 m_H^2$),
The exploration of the magnetic field dependency of the transition probability, 
along with the measurements of muon and electron $g-2$, 
can offer valuable insights into the parameter space of the inert doublet model.

\begin{figure}
\center
\begin{tabular}{ll}
\includegraphics[width=7cm]{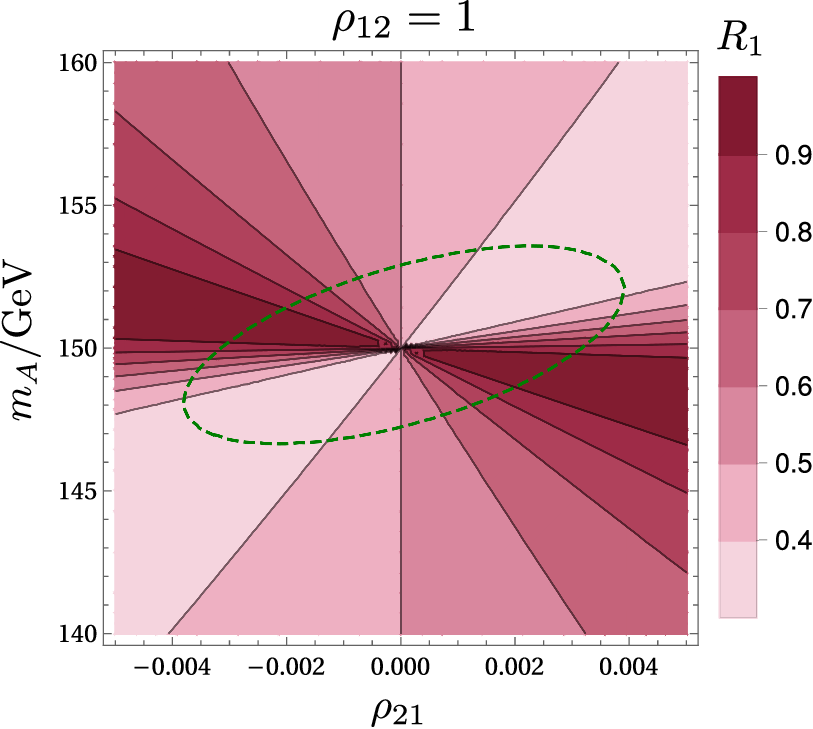}
&
\includegraphics[width=7cm]{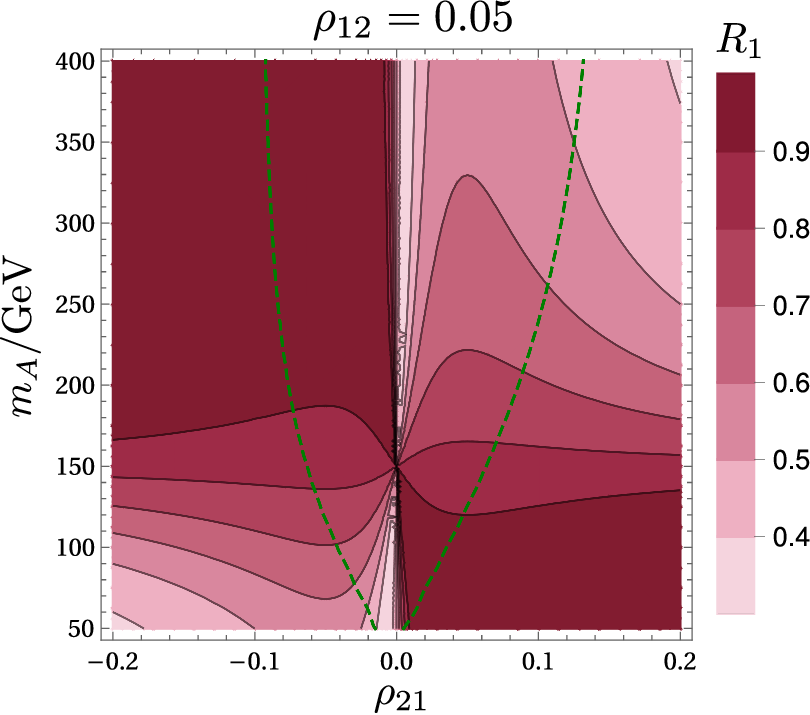} 
\\
\\
\includegraphics[width=8cm]{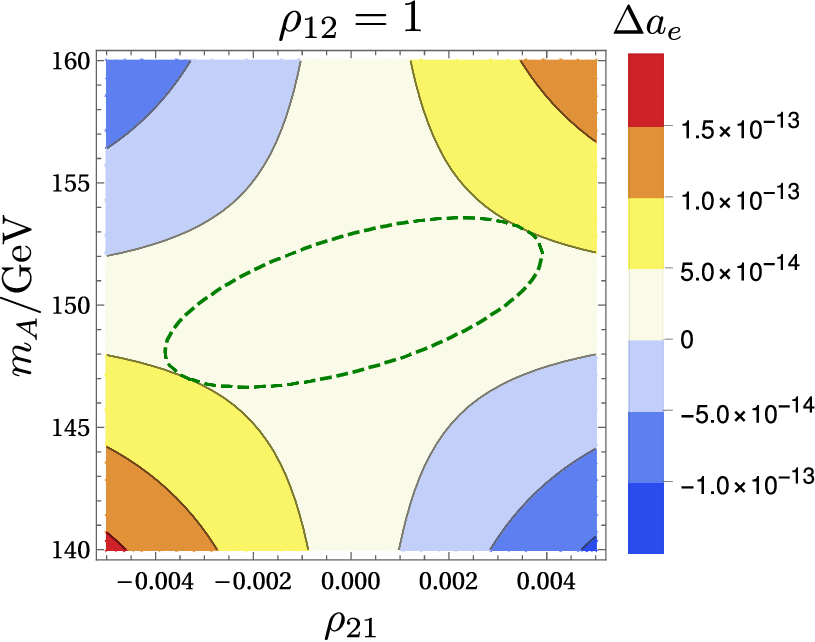}
&
\includegraphics[width=8cm]{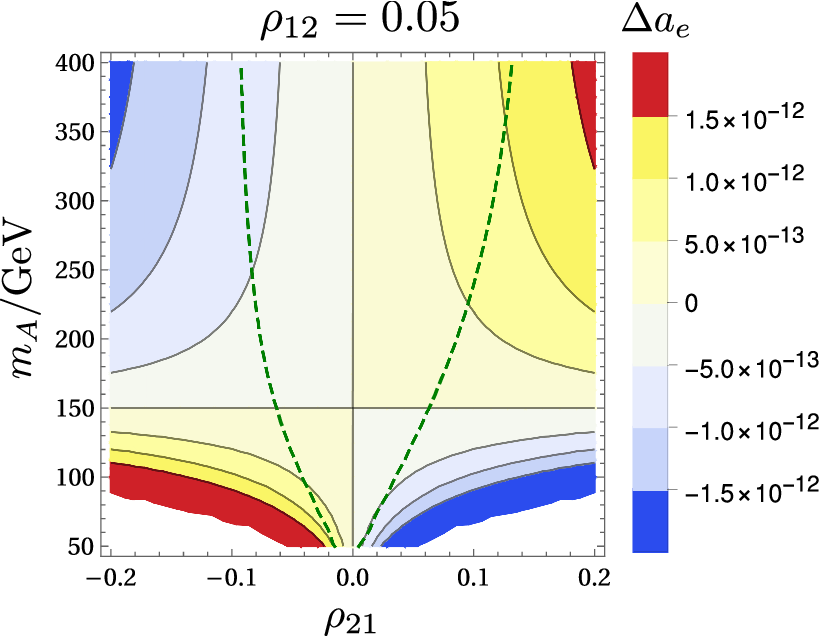}
\end{tabular}
\caption{
Contour plots illustrating the ratio of the transition probabilities (upper two plots) and 
the contribution to the electron $g-2$ (lower two plots) for 
a fixed mass of
$m_H = 150\,$GeV and varying values of $\rho_{12}$.
The green dashed lines represent contours corresponding to $\bar P = 8.3 \times 10^{-11}$,
which is the bound of the Mu-to-$\overline{\rm Mu}$ transition set by the PSI experiment.
}
\label{fig6}
\end{figure}

Fig.\ref{fig6} displays
the ratios $R_1$ and $\Delta a_e$ for $m_H = 150\,$GeV.
The green dashed lines represents the bound from the PSI experiment.
For $\rho_{12} = 1$ (left two plots),
the need for $m_A \sim m_H$ and a small $|\rho_{21}|$
arises to meet the PSI bound.
In this case, the contribution to the electron $g-2$ remains small due to $m_A \sim m_H$.
For $\rho_{12} = 0.05$ (right two plots),
a significant magnitude of $\Delta a_e$ becomes feasible.
This case showcases that
the ratio $R_1$ corresponds to a larger (smaller) value when
$\Delta a_e$ is negative (positive), aligning with expectations.

We comment that
the mass splitting among $H$, $A$ and $H^+$
can give rise to oblique corrections radiatively,
and the potential explanation of the CDF $W$ boson mass anomaly could be associated with
 this model \cite{Afik:2023vyl,Heeck:2023iqc,Bahl:2022xzi,Lee:2022gyf}.
This paper primarily focuses on the Mu-to-$\overline{\rm Mu}$ transition, 
considering this model as one option to generate the transition with a non-trivial $R_1$.
The investigation of the signals at the large hadron collider (LHC) will be explored in other works like
Ref.\cite{Afik:2023vyl}.

\section{Neutral flavor gauge boson}
\label{sec7}

The LFV neutral gauge boson couplings to induce the Mu-to-$\overline{\rm Mu}$ transition are discussed in Ref.\cite{Fukuyama:2021iyw}.
Assuming the absence of $\Delta L_e = -\Delta L_\mu = \pm 1$ processes,
we consider the following Lagrangian for the LFV neutral gauge boson (for more details, refer to Ref.\cite{Fukuyama:2022fwi}):
\begin{equation}
{\cal L} = g_X ( \overline{\ell_\mu} \gamma_\alpha \ell_e + \overline{\ell_e} \gamma_\alpha \ell_\mu) X^\alpha
+ a_X g_X ( e^{-i \varphi_X} \overline{\mu_R} \gamma_\alpha e_R + e^{i \varphi_X} \overline{e_R}\gamma_\alpha \mu_R)X^\alpha + \frac12 M_X^2 X^\alpha X_\alpha.
\end{equation}

The coefficients of the induced Mu-to-$\overline{\rm Mu}$ transition operators
are
\begin{equation}
G_1 = \frac{g_X^2}{4\sqrt2 M_X^2},
\qquad
G_2 = \frac{a_X^2 e^{-2i\varphi_X} g_X^2}{4\sqrt2 M_X^2},
\qquad
G_3 = \frac{a_X e^{-i\varphi_X} g_X^2}{2\sqrt2 M_X^2}.
\end{equation}
While a general phase $\varphi_X$ might exist,
it is constrained by the electron EDM,
and we assume $\varphi_X = 0$.
It means that there is no physical phase in the transition amplitudes due to the electron EDM.
The muon and electron $g-2$ are presented in Ref.\cite{Fukuyama:2021iyw}.
The contribution to the muon $g-2$ is negative, and thus, it cannot account for the deviation of the 
muon $g-2$.
Due to the bound of the Mu-to-$\overline{\rm Mu}$ transition,
$|\Delta a_\mu|$ is small ($\alt O(10^{-11})$).
The contribution to the electron $g-2$ is
\begin{equation}
\Delta a_e = \frac{a_X g_X^2}{16\pi^2} \frac{m_e m_\mu}{M_X^2} g\!\left(\frac{m_\mu^2}{M_X^2} \right),
\end{equation}
where $g(x)$ is a loop function, $g(x) = (4-3x-x^3+6x\ln x)/(1-x)^3$,
and the value of $g(x)$ is positive, $2< g(x)< 4$, for $0< x < 1$.
The loop function lacks a log enhancement, which distinguishes it from the scalar loop in Eq.(\ref{eq:log-enhance-scalar}).
As a result, we obtain
$|\Delta a_e| \alt O(10^{-13})$ \cite{Fukuyama:2021iyw}.

The ratio of the transition probability $R_1$ is calculated as
\begin{equation}
R_1 = \frac12 - \frac{2a_X (a_X^2 - a_X+1)}{(1+a_X^2)^2 + 3 a_X^2}.
\end{equation}
One finds 
$R_1 > 1/2$ if $\Delta a_e < 0$ and
$R_1 < 1/2$ if $\Delta a_e > 0$,
though it may be challenging to determine the sign of $\Delta a_e$ 
for $|\Delta a_e| \alt O(10^{-13})$
due to the experimental uncertainties.

We remark that 
the interaction of the neutral gauge boson can generate
a muon decay operator, $g_{RR}^S (\overline{e_R} \nu_e) (\overline{\nu_\mu} \mu_R)$,
which can interfere with the standard muon decay operator via the $W$ boson.
As a result, transverse positron polarization is induced in the decay of $\mu^+$.
In this model, the parameter $\beta$ which represents the transverse positron polarization 
is approximately proportional 
to the parameter $a_X$, as discussed in Ref.\cite{Fukuyama:2022fwi}.
If the Mu-to-$\overline{\rm Mu}$ transition is detected experimentally,
the validity of this model can be assessed through the measurements of $\beta$ and the ratio $R_1$.

\section{Conclusion}
\label{sec8}

After a quarter of a century of silence, the Mu-to-$\overline{\rm Mu}$ transition experiment
is on the verge of an update.
The MACE experiment aims to measure the time-integrated transition probability 
at a magnetic field $B_2 = 0.1\,$T.
The experiment planned at J-PARC
aims to measure the time-dependent probability
at weak magnetic fields of $B = B_0 \ll 1\,\mu$T and at a medium magnetic field of 
$B = B_1 \sim O(10) \,\mu$T,
which is approximately equal to the geomagnetic field strength.
In the $1s$ state of Mu,
there exist four states: $(F,m) = (1,\pm 1), (1,0), (0,0)$.
In a weak magnetic field,
all states, including those with $m= \pm 1$, can be involved in the transition.
However, in the magnetic field $B_1$,
the $m=\pm 1$ states are not involved in the transition
due to the energy gap generated by the magnetic field
between Mu and $\overline{\rm Mu}$.
As the magnetic field increases to $B \agt 0.01\,$T,
the $m=0$ states begin to mix, leading to the magnetic field dependence of the transition amplitudes.

The measurements of the transition probabilities at the three magnetic fields
can provide us the following information:
There are two model-independent parameters and a physical phase in the transition amplitude.
These two parameters and the phase can, in principle, be determined if
the muon polarization in the produced Mu is measured experimentally.
The physical phase should be minuscule due to the electron EDM if the transition 
is induced by a single mediator.
This allows us to ascertain the origin of the transition operator.
The ratio $R_1$ of the probability between $B=B_0$ and $B_1$
is $R_1 = 1/2$ if the mediator is a doubly charged scalar,
and $R_1 = 5/6$ if the mediator is a doubly charged gauge boson.
If the mediator is a neutral particle,
the ratio $R_1$ falls within the range of $1/3$ to $1$.
The electron $g-2$ induced by the neutral mediator, $\Delta a_e$,
is linked to the ratio $R_1$.
We find $\Delta a_e < 0$ for a larger value of $R_1$, 
and $\Delta a_e > 0$ for a smaller value of $R_1$.
Refer to Eqs.(\ref{eq:e-gminus2-Cs}) and (\ref{eq:e-gminus2-Rb})
for the current status of the electron $g-2$.
The magnitude of $\Delta a_e$ is constrained by the Mu-to-$\overline{\rm Mu}$ transition.
If the mediator is a neutral scalar,
the current experimental bound of the transition allows for 
a significant value of $|\Delta a_e|$.

We have investigated three models
for the neutral mediators: 
(1) Axion-like particle, (2) Inert doublet model, and (3) Neutral flavor gauge boson.
In model (1), it is clear how the probability ratio $R_1$ and $\Delta a_e$ are linked.
In model (2), only one of the contributions to electron and muon $g-2$ can be sizable,
satisfying the bound of the Mu-to-$\overline{\rm Mu}$ transition.
In model (3), 
the $g-2$ contributions are small due to the transition bound.
However,
a new muon decay operator, $g_{RR}^S (\overline{e_R} \nu_e) (\overline{\nu_\mu} \mu_R)$,
is induced, and the transverse positron polarization in the polarized $\mu^+$ decay
is related to the probability ratio $R_1$.
The facilities with upgraded muon beamlines
can also measure the 
the transverse positron polarization. 
It should be noted that $g_{RR}^V (\overline{e_R} \gamma_\alpha \nu_e) (\overline{\nu_\mu} \gamma^\alpha \mu_R)$
is induced in model (2) if the neutrinos are Majorana.
Unlike model (3), the $g_{RR}^V$ operator does not directly interfere with the SM amplitude.
For more details, refer to Ref.\cite{Fukuyama:2022fwi}.
By collecting data on the transition probabilities from the J-PARC and MACE experiments, 
ample insights into the origin of the Mu-to-$\overline{\rm Mu}$ transition can be gained.

\section*{Acknowledgements}

We would like to thank N.~Kawamura for the fruitful discussions.
This work was supported in part by 
JSPS KAKENHI Grant Numbers JP22H01237 (T.F. and Y.U), JP22K03602, and JP23K13106 (Y.U.).

\appendix

\section{Mu states in the magnetic field}
\label{appendix:A}

In this appendix, we provide a brief summary of the spin and energy eigenstates of Mu 
in the presence of a magnetic field.

The spin operator acts on the spin states as
\begin{equation}
S_z |\! \uparrow \rangle = \frac12|\! \uparrow \rangle , \qquad S_z |\! \downarrow \rangle = -\frac12|\! \downarrow \rangle,
\end{equation}
\begin{equation}
S_+ |\! \uparrow \rangle = 0, \qquad S_- |\! \downarrow \rangle = 0, 
\end{equation}
\begin{equation}
S_+ |\! \downarrow \rangle =  |\! \uparrow \rangle, \qquad S^- |\! \uparrow \rangle =   |\! \downarrow \rangle.
\end{equation}
The up arrow represents the state of spin $1/2$, and the down arrow represents the state of spin $-1/2$.
We note ${\bm S}^2 = S_z^2 + \frac12 (S_+ S_- + S_- S_+) = \frac34 {\bf 1}$.

We denote the spins of Mu in the order of the muon $(\mu^+)$ and the electron ($e^-$).
For example, $|{\rm Mu}; \uparrow \downarrow \rangle$ represents the spin configuration
where the muon has a spin of  $1/2$ and the electron has a spin of $-1/2$.
Noting ${\bm S}_\mu \cdot {\bm S}_e = S_\mu^z S_e^z + \frac12 (S_\mu^+ S_e^- + S_\mu^- S_e^+)$, 
we obtain
\begin{align}
&
{\bm S}_\mu \cdot {\bm S}_e |{\rm Mu}; \uparrow \uparrow \rangle = \frac14   |{\rm Mu}; \uparrow \uparrow \rangle,
\qquad
{\bm S}_\mu \cdot {\bm S}_e |{\rm Mu}; \downarrow \downarrow \rangle = \frac14   |{\rm Mu}; \downarrow \downarrow \rangle, \\
& {\bm S}_\mu \cdot {\bm S}_e |{\rm Mu};\uparrow \downarrow \rangle = -\frac14   |{\rm Mu}; \uparrow \downarrow \rangle
+ \frac12 |{\rm Mu}; \downarrow \uparrow \rangle, \\
&
{\bm S}_\mu \cdot {\bm S}_e |{\rm Mu};\downarrow \uparrow \rangle = -\frac14   |{\rm Mu};\downarrow \uparrow \rangle
+ \frac12 |{\rm Mu}; \uparrow \downarrow \rangle.
\end{align}
The eigenstates of ${\bm S}_\mu \cdot {\bm S}_e$ are
\begin{align}
&|{\rm Mu};1,1 \rangle = |{\rm Mu};\uparrow \uparrow \rangle, \qquad
|{\rm Mu};1,-1 \rangle = |{\rm Mu};\downarrow \downarrow \rangle,  \\
&|{\rm Mu};1,0 \rangle = \frac1{\sqrt2} (|{\rm Mu}; \uparrow \downarrow \rangle + |{\rm Mu}; \downarrow \uparrow \rangle ),\\
&|{\rm Mu}; 0,0 \rangle = \frac1{\sqrt2} (|{\rm Mu}; \uparrow \downarrow \rangle - |{\rm Mu}; \downarrow \uparrow \rangle ).
\end{align}
One can find
\begin{eqnarray}
{\bm S}_\mu \cdot {\bm S}_e  |{\rm Mu};1,0 \rangle = \left(-\frac14 + \frac12 \right) |{\rm Mu}; 1,0 \rangle, \\
{\bm S}_\mu \cdot {\bm S}_e  |{\rm Mu};0,0 \rangle = \left(-\frac14 - \frac12 \right) |{\rm Mu}; 0,0 \rangle.
\end{eqnarray}
The eigenvalues of ${\bm S}_\mu \cdot {\bm S}_e$ are
$1/4$ for $|{\rm Mu};1,m \rangle$ ($m = 1,0,-1$) and $-3/4$ for $|{\rm Mu}; 0, 0 \rangle$.

Now let us consider the spin Hamiltonian in the presence of a magnetic field $\bm{B}$:
\begin{equation}
{\cal H}_S = a_{\rm HFS}\, {\bm S}_\mu \cdot {\bm S}_e - {\bm \mu}_{e^-} \cdot {\bm B} - {\bm \mu}_{\mu^+} \cdot {\bm B},
\end{equation}
where $a_{\rm HFS}$ is a hyperfine structure coupling constant, 
and
${\bm \mu}_{e^-}$ and  ${\bm \mu}_{\mu^+}$ 
are the magnetic moments of the electron and muon:
\begin{equation}
{\bm \mu}_{e^-} = - g_e \mu_B {\bm S}_e,
\qquad
{\bm \mu}_{\mu^+} =  g_\mu \frac{m_e}{m_\mu} \mu_B {\bm S}_\mu.
\end{equation}
In these equations, $g_e$ and $g_\mu$ are the $g$-factors of the electron and muon,
and $\mu_B$ is the Bohr magneton.
We define two dimensionless quantities as follows:
\begin{align}
X &= \frac1{a_{\rm HFS}} \mu_B B (g_e + \frac{m_e}{m_\mu} g_\mu) \simeq 6.31 B/{\rm Tesla},  \label{eq:X}\\
Y &= \frac1{a_{\rm HFS}} \mu_B B (g_e - \frac{m_e}{m_\mu} g_\mu) \simeq 6.25 B/{\rm Tesla}. 
\end{align}
Then, 
supposing that the magnetic field is aligned with the $z$-direction, 
we obtain
\begin{align}
{\cal H}_S |{\rm Mu};1,\pm 1 \rangle &= a_{\rm HFS} \left(\frac{1}4 \pm \frac{Y}{2} \right) |{\rm Mu};1,\pm 1 \rangle, \\
{\cal H}_S |{\rm Mu};1,0 \rangle & = a_{\rm HFS} \left( \frac{1}4|{\rm Mu};1,0 \rangle - \frac{X}2 | {\rm Mu};0,0 \rangle \right), \\
{\cal H}_S |{\rm Mu};0,0 \rangle & = a_{\rm HFS} \left( -\frac{3}4|{\rm Mu};1,0 \rangle - \frac{X}2 | {\rm Mu};1,0 \rangle \right).
\end{align}
Consequently, for $B\neq 0$, $|{\rm Mu}; 1,0 \rangle$ and $|{\rm Mu}; 0,0 \rangle$ are not energy eigenstates.
The energy eigenstates in a magnetic field are given as
\begin{equation}
\left(
\begin{array}{c} 
|{\rm Mu};1,0 \rangle_B \\
|{\rm Mu};0,0 \rangle_B 
\end{array}  
\right)
=
\left(
\begin{array}{cc}
C & - S \\
S & C 
\end{array}
\right)
\left(
\begin{array}{c} 
|{\rm Mu};1,0 \rangle \\
|{\rm Mu};0,0 \rangle 
\end{array}  
\right) ,
\end{equation}
where $C = \cos (\frac12 \arctan X)$ and $S = \sin(\frac12 \arctan X)$:
\begin{equation}
C^2 = \frac12 \left(1 + \frac1{\sqrt{1+X^2}} \right),
\qquad
S^2 = \frac12 \left(1 - \frac1{\sqrt{1+X^2}} \right).
\end{equation}
We obtain
\begin{align}
{\cal H}_S |{\rm Mu}; 1,0 \rangle_B &= a_{\rm HFS} \left( - \frac14 + \frac12 \sqrt{1+X^2} \right)|{\rm Mu}; 1,0 \rangle_B,\\
{\cal H}_S |{\rm Mu}; 0,0 \rangle_B &= a_{\rm HFS} \left( - \frac14 - \frac12 \sqrt{1+X^2} \right) |{\rm Mu}; 0,0 \rangle_B.
\end{align}
It is convenient to give the states as
\begin{align}
\left(
\begin{array}{c} 
 |{\rm Mu}; 1,0 \rangle_B \\
|{\rm Mu};0,0 \rangle_B 
\end{array}  
\right)
=
 \left(
\begin{array}{cc}
c & s \\
-s & c 
\end{array}
\right)
\left(
\begin{array}{c} 
|{\rm Mu}; \downarrow \uparrow \rangle \\
|{\rm Mu}; \uparrow \downarrow \rangle 
\end{array}  
\right) ,
\label{eq:Mu-in-B}
\end{align}
where $c = (C+S)/\sqrt2$ and $s = (C-S)/\sqrt2$:
\begin{equation}
c^2  = \frac12 \left( 1+ \frac{X}{\sqrt{1+X^2}} \right),
\qquad
s^2  = \frac12 \left( 1- \frac{X}{\sqrt{1+X^2}} \right).
\label{eq:s-c}
\end{equation}
In the limit of a strong magnetic field ($X \to \infty$),
the following approximations hold:
\begin{equation}
| {\rm Mu};1, 0 \rangle_B \approx |{\rm Mu};\downarrow \uparrow \rangle,
\qquad
|{\rm Mu}; 0, 0 \rangle_B \approx|{\rm Mu}; \uparrow \downarrow \rangle .
\end{equation}
This is a physically reasonable observation since the electron magnetic moment dominantly influences the states 
in strong magnetic fields, making the hyperfine structure coupling relatively negligible.
For $\overline{\rm Mu}$, where the directions of the muon and electron magnetic moments are opposite,
we have
\begin{align}
\left(
\begin{array}{c} 
 |\overline{\rm Mu}; 1,0 \rangle_B \\
|\overline{\rm Mu};0,0 \rangle_B 
\end{array}  
\right)
=
 \left(
\begin{array}{cc}
s & c \\
-c & s 
\end{array}
\right)
\left(
\begin{array}{c} 
|\overline{\rm Mu}; \downarrow \uparrow \rangle \\
|\overline{\rm Mu}; \uparrow \downarrow \rangle 
\end{array}  
\right) ,
\label{eq:Mubar-in-B}
\end{align}
and in the limit of a strong magnetic field, we obtain
\begin{equation}
| \overline{\rm Mu};1, 0 \rangle_B \approx |\overline{\rm Mu}; \uparrow \downarrow \rangle,
\qquad
|\overline{\rm Mu}; 0, 0 \rangle_B \approx |\overline{\rm Mu}; \downarrow \uparrow \rangle .
\end{equation}
It should be noted that the muon and electron spins must flip 
in order for the Mu-to-$\overline{\rm Mu}$ transition to occur in the limit of a strong magnetic field.

\section{Population of the states and muon polarization in the magnetic field}
\label{appendix:B}

In this Appendix, we outline how the determination of 
the populations of produced Mu states $(F,m)$, denoted as $f_{F,m}$, is
undertaken.
We assume that the electrons are unpolarized.
Under this assumption, the populations of the states,
$|{\uparrow\uparrow} \rangle , |{\uparrow\downarrow}\rangle,
|{\downarrow\uparrow}\rangle,  |{\downarrow\downarrow}\rangle$,
where the up and down direction arrows indicate the spins of Mu
in the ordering of the muon and electron,
satisfy
\begin{equation}
f_{\uparrow\uparrow} : f_{\uparrow\downarrow}: f_{\downarrow\uparrow}: f_{\downarrow\downarrow}=
a:a:b:b.
\end{equation}
This relation results in the following connections:
\begin{equation}
f_{1,1} + f_{1,-1} = f_{1,0} + f_{0,0} = \frac12.
\end{equation}
We assume that the angular momentum transfer is negligible in the potential $2p \to 1s$ transition at the time of Mu formation.
The populations are parameterized as
\begin{equation}
(f_{\uparrow\uparrow} , f_{\uparrow\downarrow}, f_{\downarrow\uparrow}, f_{\downarrow\downarrow})=
\left( \frac{1+P_\mu}4, \frac{1+P_\mu}4, \frac{1-P_\mu}4, \frac{1-P_\mu}4 \right),
\label{eq:Pmu}
\end{equation}
where $P_\mu$ can be identified to the muon polarization in the produced Mu in the direction of the magnetic field.
We then obtain the expressions:
\begin{align}
f_{1,0} &= c^2 \frac{1-P_\mu}4 + s^2 \frac{1+P_\mu}4 =  \frac14 \left( 1 - P_\mu \frac{X}{\sqrt{1+X^2}} \right), \label{eq:f10} \\
f_{0,0} &= s^2 \frac{1+P_\mu}4 + c^2 \frac{1+P_\mu}4 = \frac14 \left( 1 + P_\mu \frac{X}{\sqrt{1+X^2}} \right), \label{eq:f00}
\end{align}
where $s$ and $c$ are given in Eq.(\ref{eq:s-c}).

We note that
the transition probability does not depend on $P_\mu$
for the regimes $X\ll 1$ (where $f_{1,0} \simeq f_{0,0} \simeq 1/4$)
and 
$X\gg 1$ (where $|{\cal M}_{1,0}| \simeq |{\cal M}_{0,0}|$).
The transition probability exhibits significant dependence on $P_\mu$
for $O(10)\, {\rm mT} < B < O(1)\, {\rm T}$.

Because 
$|{\uparrow\downarrow}\rangle$ and
$|{\downarrow\uparrow}\rangle$
are not energy eigenstates,
they can oscillate between each other 
with an oscillation time determined by the hyperfine splitting, whose time scale is approximately $0.2\,$ns. 
The muon polarization of these states is effectively averaged out and eliminated
over a time about $0.1\,\mu$s in a weak magnetic field ($<O(1)\,$mT).
As a result, the total muon polarization becomes halved.
We remark that 
the parameter $P_\mu$, defined in Eq.(\ref{eq:Pmu}),
represents the muon polarization before the averaging-out process.

It is important to mention that in experiments using noble gases as targets to produce Mu, 
the initial polarization of muon beam can be maintained.
However, the Mu-to-$\overline{\rm Mu}$ transitions are suppressed in the gases \cite{Feinberg:1961zza}.
The transition experiments will employ SiO$_2$ targets to produce Mu in a vacuum environment,
and the initial muon polarization will experience partial loss during the production of Mu.

\section{Transition amplitudes of the spin eigenstates}
\label{appendix:C}

We reexamine the  Mu-to-$\overline{\rm Mu}$ transition amplitudes
from the transition operators.
For more detailed calculation of the transition amplitudes
refer to Appendix in Ref.\cite{Fukuyama:2021iyw}.
In this Appendix, we provide the expressions of the transition amplitudes
in the spin eigenstates of Mu and $\overline{\rm Mu}$.
These expressions are useful to understand the magnetic field dependence on the amplitudes 
across different operators.

The four-component spinors for Dirac particle and anti-particle under a non-relativistic limit are expressed as
\begin{equation}
u_l (s) = \sqrt{m_l} \left( 
 \begin{array}{c}
  \xi^{s} \\ \xi^{s}
 \end{array}
\right), \qquad
v_l (s) = \sqrt{m_l} \left( 
 \begin{array}{c}
  \eta^{s} \\ -\eta^{s}
 \end{array}
\right),
\end{equation}
where $ l = e,\mu$.
The two-component spinors $\xi^s$ and $\eta^s$ are given in the spin eigenstates as follows:
$\xi^{+1/2} = (1,0)^T$,  
$\xi^{-1/2} = (0,1)^T$,
$\eta^{+1/2} = (0,1)^T$,  
and $\eta^{-1/2} = (-1,0)^T$.
The transition amplitudes in the spin eigenstates via the $S\times S$ and $P \times P$ operators,
which are given in Eqs.(\ref{eq:QS}) and (\ref{eq:QP}), respectively, 
can be obtained as
\begin{align}
\langle \overline{\rm Mu}; \bar s_\mu, \bar s_e | Q_S | {\rm Mu}; s_\mu, s_e \rangle
&=
2 |\varphi(0)|^2 (\xi^{\bar s_\mu}{}^\dagger \xi^{s_e}) (\eta^{s_\mu}{}^\dagger \eta^{\bar s_e}),  \label{eq:MQS}\\
\langle \overline{\rm Mu}; \bar s_\mu, \bar s_e | Q_P | {\rm Mu}; s_\mu, s_e \rangle
&=
-2 |\varphi(0)|^2 (\xi^{\bar s_\mu}{}^\dagger \eta^{\bar s_e}) (\eta^{s_\mu}{}^\dagger \xi^{s_e}). \label{eq:MQP}
\end{align}
Remind that 
any four-fermion operators for the Mu-to-$\overline{\rm Mu}$ transition
can be written as a linear combination of the operators $Q_i$ 
given in Eqs.(\ref{eq:Q1})\dash(\ref{eq:Q5}) through the application of Fierz transformation.
It is worth mentioning that
the transition amplitudes in the spin eigenstates through any operators
 under the non-relativistic limit
can be expressed as a linear combination of the right-hand sides of Eqs.(\ref{eq:MQS}) and (\ref{eq:MQP}).
This is similarly to the way
any $2\times 2$ matrix is expanded using a complete orthogonal system (${\bf 1}_{2\times 2}, \sigma_i$),
which is analogous to the Fierz transformation for four-component spinors.
Specifically, the amplitudes for $Q_4$ and $Q_5$ sandwiched between $\langle \overline{\rm Mu}|$
and $|{\rm Mu} \rangle$ are the same in the non-relativistic limit.
At the operator level, 
the relationships hold:
\begin{equation}
Q_4 + Q_5 = 2 (Q_S + Q_P),
\qquad
Q_3 = 2 (Q_P - Q_S).
\end{equation}
Thus, the above statement is trivial for $Q_3$, $Q_4$ and $Q_5$.
Let us examine the $V\times V$ and $A\times A$ operators:
\begin{equation}
Q_V \equiv (\bar \mu \gamma_\alpha e) (\bar \mu \gamma^\alpha e),
\qquad
Q_A \equiv (\bar \mu \gamma_\alpha \gamma_5 e) (\bar \mu \gamma^\alpha \gamma_5 e).
\end{equation}
The transition amplitude in the spin eigenstates via the $V\times V$ operator is obtained as 
\begin{equation}
\langle \overline{\rm Mu}; \bar s_\mu, \bar s_e | Q_V | {\rm Mu}; s_\mu, s_e \rangle
=
2 |\varphi(0)|^2 (-(\xi^{\bar s_\mu}{}^\dagger \sigma_i \eta^{\bar s_e}) (\eta^{s_\mu}{}^\dagger \sigma_i \xi^{s_e})
-
(\xi^{\bar s_\mu}{}^\dagger \xi^{s_e}) (\eta^{s_\mu}{}^\dagger \eta^{\bar s_e})).
\label{eq:MQV}
\end{equation}
Utilizing the complete orthogonal system in the $2\times 2$ matrices, 
one can derive the relation,
\begin{equation}
(\xi^{\bar s_\mu}{}^\dagger \sigma_i \eta^{\bar s_e}) (\eta^{s_\mu}{}^\dagger \sigma_i \xi^{s_e})
=
2 (\xi^{\bar s_\mu}{}^\dagger \xi^{s_e}) (\eta^{s_\mu}{}^\dagger \eta^{\bar s_e})
-
(\xi^{\bar s_\mu}{}^\dagger \eta^{\bar s_e}) (\eta^{s_\mu}{}^\dagger \xi^{s_e}),
\label{eq:3-vector-formula}
\end{equation}
leading to the following expression for the transition amplitude:
\begin{equation}
\langle \overline{\rm Mu}; \bar s_\mu, \bar s_e | Q_V | {\rm Mu}; s_\mu, s_e \rangle
=
2 |\varphi(0)|^2 (
-3
(\xi^{\bar s_\mu}{}^\dagger \xi^{s_e}) (\eta^{s_\mu}{}^\dagger \eta^{\bar s_e})
+
(\xi^{\bar s_\mu}{}^\dagger \eta^{\bar s_e}) (\eta^{s_\mu}{}^\dagger \xi^{s_e})) .
\end{equation}
One can also find the amplitude via the $A\times A$ operator:
\begin{equation}
\langle \overline{\rm Mu}; \bar s_\mu, \bar s_e | Q_A | {\rm Mu}; s_\mu, s_e \rangle
=
2 |\varphi(0)|^2 (
-
(\xi^{\bar s_\mu}{}^\dagger \xi^{s_e}) (\eta^{s_\mu}{}^\dagger \eta^{\bar s_e})
+3
(\xi^{\bar s_\mu}{}^\dagger \eta^{\bar s_e}) (\eta^{s_\mu}{}^\dagger \xi^{s_e})) .
\end{equation}
Similarly to the case of $Q_4$ and $Q_5$,
the amplitudes for $Q_1$ and $Q_2$ sandwiched between $\langle \overline{\rm Mu}|$
and $|{\rm Mu} \rangle$ are the same in the non-relativistic limit,
and $Q_1+ Q_2 = 2(Q_V + Q_A) $ holds.
These allow us to derive the following relationship:
\begin{align}
& \langle \overline{\rm Mu}; \bar s_\mu, \bar s_e | Q_1 | {\rm Mu}; s_\mu, s_e \rangle
=
\langle \overline{\rm Mu}; \bar s_\mu, \bar s_e | Q_2 | {\rm Mu}; s_\mu, s_e \rangle \nonumber \\
&=
-4 \langle \overline{\rm Mu}; \bar s_\mu, \bar s_e | Q_4 | {\rm Mu}; s_\mu, s_e \rangle
=
-4 \langle \overline{\rm Mu}; \bar s_\mu, \bar s_e | Q_5 | {\rm Mu}; s_\mu, s_e \rangle \nonumber \\
& =
- 8
 |\varphi(0)|^2 \left(
(\xi^{\bar s_\mu}{}^\dagger \xi^{s_e}) (\eta^{s_\mu}{}^\dagger \eta^{\bar s_e})
-
(\xi^{\bar s_\mu}{}^\dagger \eta^{\bar s_e}) (\eta^{s_\mu}{}^\dagger \xi^{s_e}) \right) .
\label{eq:Q1245xieta}
\end{align}
For $Q_3$, we obtain from $Q_3 = 2(Q_P-Q_S)$,
\begin{align}
 \langle \overline{\rm Mu}; \bar s_\mu, \bar s_e | Q_3 | {\rm Mu}; s_\mu, s_e \rangle
 =
-2  |\varphi(0)|^2 \left(
(\xi^{\bar s_\mu}{}^\dagger \xi^{s_e}) (\eta^{s_\mu}{}^\dagger \eta^{\bar s_e})
+
(\xi^{\bar s_\mu}{}^\dagger \eta^{\bar s_e}) (\eta^{s_\mu}{}^\dagger \xi^{s_e}) \right).
\label{eq:Q3xieta}
\end{align}
Thus, the above statement is presented.
As a result, the transition amplitude under the non-relativistic limit can be expressed as a function of two degrees of freedom.

With these equations at hand,
we can incidentally reproduce Eqs.(\ref{eq:M1m}) and (\ref{eq:M00}).
To provide clarify, we will use up and down direction arrows to symbolize the spins of Mu and $\overline{\rm Mu}$
in the ordering of muons and electrons,
as described in Appendix \ref{appendix:A}.
 In the case of $Q_S$, one obtains
\begin{align}
&\langle \overline{\rm Mu} ; \uparrow \uparrow | Q_S | {\rm Mu} ; \uparrow \uparrow \rangle
=
\langle \overline{\rm Mu} ; \downarrow \downarrow | Q_S | {\rm Mu} ; \downarrow \downarrow \rangle \nonumber \\
& = 
\langle \overline{\rm Mu} ; \downarrow \uparrow | Q_S | {\rm Mu} ; \uparrow \downarrow \rangle
=
\langle \overline{\rm Mu} ; \uparrow \downarrow | Q_S | {\rm Mu} ; \downarrow \uparrow \rangle 
= 2|\varphi(0)|^2,
\label{eq:app-Qs-spin}
\end{align}
and the amplitudes for the other spin combinations vanish.
Then, the amplitudes in the energy eigenstates 
are calculated as
\begin{equation}
\langle \overline{\rm Mu} ; 1,m | Q_S | {\rm Mu} ; 1,m \rangle
= 2|\varphi(0)|^2,
\quad
\langle \overline{\rm Mu} ; 0,0 | Q_S | {\rm Mu} ; 0,0 \rangle
= -2|\varphi(0)|^2.
\label{eq:amp-QS}
\end{equation}
In the case of $Q_P$,
one obtains
\begin{align}
&\langle \overline{\rm Mu} ; \uparrow \downarrow | Q_P | {\rm Mu} ; \uparrow \downarrow \rangle
=
\langle \overline{\rm Mu} ; \downarrow \uparrow | Q_P | {\rm Mu} ; \downarrow \uparrow \rangle =
2|\varphi(0)|^2, \\
& 
\langle \overline{\rm Mu} ; \downarrow \uparrow | Q_P | {\rm Mu} ; \uparrow \downarrow \rangle
=
\langle \overline{\rm Mu} ; \uparrow \downarrow | Q_P | {\rm Mu} ; \downarrow \uparrow \rangle 
= -2|\varphi(0)|^2,
\end{align}
and the amplitudes for the other spin combinations vanish.
This gives the following amplitudes in the energy eigenstates:
\begin{equation}
\langle \overline{\rm Mu} ; 1,m | Q_P | {\rm Mu} ; 1,m \rangle
= 0,
\quad
\langle \overline{\rm Mu} ; 0,0 | Q_P | {\rm Mu} ; 0,0 \rangle
= 4|\varphi(0)|^2.
\label{eq:amp-QP}
\end{equation}
Utilizing Eq.(\ref{eq:Q1245xieta}), we find
\begin{align}
&\langle \overline{\rm Mu} ; F,m | Q_1 | {\rm Mu} ; F,m \rangle
=
\langle \overline{\rm Mu} ; F,m | Q_2 | {\rm Mu} ; F,m \rangle \nonumber \\
& =
-4 \langle \overline{\rm Mu} ; F,m | Q_4 | {\rm Mu} ; F,m \rangle
=
-4 \langle \overline{\rm Mu} ; F,m | Q_5 | {\rm Mu} ; F,m \rangle \nonumber \\
&=
-4 \left( \langle \overline{\rm Mu} ; F,m | Q_S | {\rm Mu} ; F,m \rangle + \langle \overline{\rm Mu} ; F,m | Q_P | {\rm Mu} ; F,m \rangle \right),
\end{align}
and with $Q_3 = 2(Q_P-Q_S)$,
\begin{align}
\langle \overline{\rm Mu} ; F,m | Q_3 | {\rm Mu} ; F,m \rangle
=
2 \left( \langle \overline{\rm Mu} ; F,m | Q_P | {\rm Mu} ; F,m \rangle - \langle \overline{\rm Mu} ; F,m | Q_S | {\rm Mu} ; F,m \rangle \right).
\end{align}
Substituting Eqs.(\ref{eq:amp-QS}) and (\ref{eq:amp-QP}),
one can reproduce Eqs.(\ref{eq:M1m}) and (\ref{eq:M00}).

We note that one can find
\begin{equation}
\langle \overline{\rm Mu} ; 1,m | Q_1 | {\rm Mu} ; 1,m \rangle = \langle \overline{\rm Mu} ; 0,0 | Q_1 | {\rm Mu} ; 0,0 \rangle,
\end{equation}
and the same for $Q_2$, $Q_4$ and $Q_5$.
The equivalence of the amplitudes for the triplet and singlet states can be readily derived from the transition amplitudes in the spin eigenstates:
\begin{align}
&\langle \overline{\rm Mu} ; \uparrow \uparrow | Q_1 | {\rm Mu} ; \uparrow \uparrow \rangle
=
\langle \overline{\rm Mu} ; \downarrow \downarrow | Q_1 | {\rm Mu} ; \downarrow \downarrow \rangle \nonumber \\
& = 
\langle \overline{\rm Mu} ; \downarrow \uparrow | Q_1 | {\rm Mu} ; \downarrow \uparrow \rangle
=
\langle \overline{\rm Mu} ; \uparrow \downarrow | Q_1 | {\rm Mu} ; \uparrow \downarrow \rangle 
= -8|\varphi(0)|^2,
\end{align}
and the amplitudes for the other spin combinations vanish,
which can be straightforwardly deduced using Eq.(\ref{eq:Q1245xieta})
and an identity equation for two-component spinors,
\begin{equation}
(\xi^{\bar s_\mu}{}^\dagger \xi^{s_e}) (\eta^{s_\mu}{}^\dagger \eta^{\bar s_e})-(\xi^{\bar s_\mu}{}^\dagger \eta^{\bar s_e}) (\eta^{s_\mu}{}^\dagger \xi^{s_e})
= (\xi^{\bar s_\mu}{}^T \epsilon \eta^{s_\mu})^* (\xi^{s_e}{}^T \epsilon \eta^{\bar s_e}),
\end{equation}
where $\epsilon = i \sigma_2$.
In connection with this, we can derive an identity equation,
\begin{equation}
(\xi^{\bar s_\mu}{}^\dagger \xi^{s_e}) (\eta^{s_\mu}{}^\dagger \eta^{\bar s_e})+(\xi^{\bar s_\mu}{}^\dagger \eta^{\bar s_e}) (\eta^{s_\mu}{}^\dagger \xi^{s_e})
= (\xi^{\bar s_\mu}{}^T \sigma_i  \epsilon \eta^{s_\mu})^* (\xi^{s_e}{}^T  \sigma_i \epsilon \eta^{\bar s_e}),
\end{equation}
pertaining to the transition amplitude through $Q_3$ in the spin eigenstate in Eq.(\ref{eq:Q3xieta}).
The right-hand sides of these two equations correspond explicitly
to the $t$-channel exchanges of the doubly charged particles,
as depicted in Fig.\ref{fig3} and explained in Section \ref{sec3}.


\begin{thebibliography}{99}


%-----
%  muon beamlines

%\cite{Kawamura:2018apy}
\bibitem{Kawamura:2018apy}
N.~Kawamura \textit{et al.}
%N.~Kawamura, M.~Aoki, J.~Doornbos, T.~Mibe, Y.~Miyake, F.~Morimoto, Y.~Nakatsugawa, M.~Otani, N.~Saito and Y.~Seiya, \textit{et al.}
``New concept for a large-acceptance general-purpose muon beamline,''
PTEP \textbf{2018}, no.11, 113G01 (2018)
\doi{10.1093/ptep/pty116}

%\cite{Aiba:2021bxe}
\bibitem{Aiba:2021bxe}
M.~Aiba \textit{et al.}
%M.~Aiba, A.~Amato, A.~Antognini, S.~Ban, N.~Berger, L.~Caminada, R.~Chislett, P.~Crivelli, A.~Crivellin and G.~D.~Maso, \textit{et al.}
``Science Case for the new High-Intensity Muon Beams HIMB at PSI,''
[arXiv:2111.05788 [hep-ex]].



%----
%
%(LFV)


%\cite{Adam:2013mnn}
\bibitem{Adam:2013mnn}
J.~Adam \textit{et al.} [MEG],
``New constraint on the existence of the $\mu^+ \to e^+\gamma$ decay,''
Phys. Rev. Lett. \textbf{110}, 201801 (2013)
\doi{10.1103/PhysRevLett.110.201801}
[arXiv:1303.0754 [hep-ex]];
%
%\cite{TheMEG:2016wtm}
%\bibitem{TheMEG:2016wtm}
A.~M.~Baldini \textit{et al.} [MEG],
``Search for the lepton flavour violating decay $\mu ^+ \rightarrow \mathrm {e}^+ \gamma $ with the full dataset of the MEG experiment,''
Eur. Phys. J. C \textbf{76}, no.8, 434 (2016)
\doi{10.1140/epjc/s10052-016-4271-x}
[arXiv:1605.05081 [hep-ex]].


%\cite{Bellgardt:1987du}
\bibitem{Bellgardt:1987du}
U.~Bellgardt \textit{et al.} [SINDRUM],
``Search for the Decay $\mu^+ \to e^+ e^+ e^-$,''
Nucl. Phys. B \textbf{299}, 1-6 (1988)
\doi{10.1016/0550-3213(88)90462-2}

%\cite{SINDRUMII:2006dvw}
\bibitem{SINDRUMII:2006dvw}
W.~H.~Bertl \textit{et al.} [SINDRUM II],
``A Search for muon to electron conversion in muonic gold,''
Eur. Phys. J. C \textbf{47}, 337-346 (2006)
\doi{10.1140/epjc/s2006-02582-x}



% ------
% Mu-to-Mubar transition
  

%\cite{Pontecorvo:1957cp}
\bibitem{Pontecorvo:1957cp}
B.~Pontecorvo,
``Mesonium and anti-mesonium,''
Sov. Phys. JETP \textbf{6}, 429 (1957)



%\cite{Feinberg:1961zza}
\bibitem{Feinberg:1961zza}
G.~Feinberg and S.~Weinberg,
``Conversion of Muonium into Antimuonium,''
Phys. Rev. \textbf{123}, 1439-1443 (1961)
\doi{10.1103/PhysRev.123.1439}


%\cite{Lee:1977tib}
\bibitem{Lee:1977tib}
B.~W.~Lee and R.~E.~Shrock,
``Natural Suppression of Symmetry Violation in Gauge Theories: Muon - Lepton and Electron Lepton Number Nonconservation,''
Phys. Rev. D \textbf{16}, 1444 (1977)
\doi{10.1103/PhysRevD.16.1444};
%
%\cite{Lee:1977qz}
%\bibitem{Lee:1977qz}
B.~W.~Lee, S.~Pakvasa, R.~E.~Shrock and H.~Sugawara,
``Muon and Electron Number Nonconservation in a V-A Gauge Model,''
Phys. Rev. Lett. \textbf{38}, 937 (1977)
[erratum: Phys. Rev. Lett. \textbf{38}, 1230 (1977)]
\doi{10.1103/PhysRevLett.38.937}


%\cite{Halprin:1982wm}
\bibitem{Halprin:1982wm}
A.~Halprin,
``Neutrinoless Double Beta Decay and Muonium - Anti-Muonium Transitions,''
Phys. Rev. Lett. \textbf{48}, 1313-1316 (1982)
\doi{10.1103/PhysRevLett.48.1313}



%--- experiments

%\cite{Willmann:1998gd}
\bibitem{Willmann:1998gd}
L.~Willmann, P.~V.~Schmidt, H.~P.~Wirtz, R.~Abela, V.~Baranov, J.~Bagaturia, W.~H.~Bertl, R.~Engfer, A.~Grossmann and V.~W.~Hughes, \textit{et al.}
``New bounds from searching for muonium to anti-muonium conversion,''
Phys. Rev. Lett. \textbf{82}, 49-52 (1999)
\doi{10.1103/PhysRevLett.82.49}
[arXiv:hep-ex/9807011 [hep-ex]].





%\cite{Han:2021nod}
\bibitem{Han:2021nod}
C.~Han, D.~Huang, J.~Tang and Y.~Zhang,
``Probing the doubly-charged Higgs with Muonium to Antimuonium Conversion Experiment,''
Phys. Rev. D \textbf{103}, no.5, 055023 (2021)
\doi{10.1103/PhysRevD.103.055023}
[arXiv:2102.00758 [hep-ph]].

%\cite{Bai:2022sxq}
\bibitem{Bai:2022sxq}
A.~Y.~Bai, Y.~Chen, Y.~Chen, R.~R.~Fan, Z.~Hou, H.~T.~Jing, H.~B.~Li, Y.~Li, H.~Miao and H.~Peng, \textit{et al.}
``Snowmass2021 Whitepaper: Muonium to antimuonium conversion,''
[arXiv:2203.11406 [hep-ph]].





%\cite{Kawamura:2021lqk}
\bibitem{Kawamura:2021lqk}
N.~Kawamura, R.~Kitamura, H.~Yasuda, M.~Otani, Y.~Nakazawa, H.~Iinuma and T.~Mibe,
``A New Approach for Mu - \(\overline{\text{Mu}}\) Conversion Search,''
JPS Conf. Proc. \textbf{33}, 011120 (2021)
\doi{10.7566/JPSCP.33.011120}



%-----------

%\cite{Conlin:2020veq}
\bibitem{Conlin:2020veq}
R.~Conlin and A.~A.~Petrov,
``Muonium-antimuonium oscillations in effective field theory,''
Phys. Rev. D \textbf{102}, no.9, 095001 (2020)
\doi{10.1103/PhysRevD.102.095001}
[arXiv:2005.10276 [hep-ph]].


%-------------


%\cite{Fukuyama:2021iyw}
\bibitem{Fukuyama:2021iyw}
T.~Fukuyama, Y.~Mimura and Y.~Uesaka,
``Models of the muonium to antimuonium transition,''
Phys. Rev. D \textbf{105}, no.1, 015026 (2022)
\doi{10.1103/PhysRevD.105.015026}
[arXiv:2108.10736 [hep-ph]].






%--------



%(Magnetic field dependence)

%\cite{Horikawa:1995ae}
\bibitem{Horikawa:1995ae}
K.~Horikawa and K.~Sasaki,
``Muonium - anti-muonium conversion in models with dilepton gauge bosons,''
Phys. Rev. D \textbf{53}, 560-563 (1996)
\doi{10.1103/PhysRevD.53.560}
[arXiv:hep-ph/9504218 [hep-ph]].


%\cite{Hou:1995np}
\bibitem{Hou:1995np}
W.~S.~Hou and G.~G.~Wong,
``Magnetic field dependence of muonium - anti-muonium conversion,''
Phys. Lett. B \textbf{357}, 145-150 (1995)
\doi{10.1016/0370-2693(95)00893-P}
[arXiv:hep-ph/9505300 [hep-ph]].



% electron EDM
%
%\cite{ACME:2018yjb}
\bibitem{ACME:2018yjb}
V.~Andreev \textit{et al.} [ACME],
``Improved limit on the electric dipole moment of the electron,''
Nature \textbf{562}, no.7727, 355-360 (2018)
\doi{10.1038/s41586-018-0599-8}

%\cite{Roussy:2022cmp}
\bibitem{Roussy:2022cmp}
T.~S.~Roussy, L.~Caldwell, T.~Wright, W.~B.~Cairncross, Y.~Shagam, K.~B.~Ng, N.~Schlossberger, S.~Y.~Park, A.~Wang and J.~Ye, \textit{et al.}
``An improved bound on the electron\textquoteright{}s electric dipole moment,''
Science \textbf{381}, no.6653, adg4084 (2023)
\doi{10.1126/science.adg4084}
[arXiv:2212.11841 [physics.atom-ph]].


%----
%
%(Doubly charged scalar)


%\cite{Chang:1989uk}
\bibitem{Chang:1989uk}
D.~Chang and W.~Y.~Keung,
``Constraints on Muonium-antiMuonium Conversion,''
Phys. Rev. Lett. \textbf{62}, 2583 (1989)
\doi{10.1103/PhysRevLett.62.2583}


%\cite{Swartz:1989qz}
\bibitem{Swartz:1989qz}
M.~L.~Swartz,
``Limits on Doubly Charged Higgs Bosons and Lepton Flavor Violation,''
Phys. Rev. D \textbf{40}, 1521 (1989)
\doi{10.1103/PhysRevD.40.1521}




%-----
%

%(dilepton gauge boson)


%\cite{Frampton:1989fu}
\bibitem{Frampton:1989fu}
P.~H.~Frampton and B.~H.~Lee,
``SU(15) GRAND UNIFICATION,''
Phys. Rev. Lett. \textbf{64}, 619 (1990)
\doi{10.1103/PhysRevLett.64.619};
%
%\cite{Frampton:1991mt}
%\bibitem{Frampton:1991mt}
P.~H.~Frampton and D.~Ng,
``Dileptons: Present status and future prospects,''
Phys. Rev. D \textbf{45}, 4240-4245 (1992)
\doi{10.1103/PhysRevD.45.4240};
%
%\cite{Frampton:1992wt}
%\bibitem{Frampton:1992wt}
P.~H.~Frampton,
``Chiral dilepton model and the flavor question,''
Phys. Rev. Lett. \textbf{69}, 2889-2891 (1992)
\doi{10.1103/PhysRevLett.69.2889}


%\cite{Fujii:1992np}
\bibitem{Fujii:1992np}
H.~Fujii, S.~Nakamua and K.~Sasaki,
``Constraints on dilepton mass from low-energy muon experiments,''
Phys. Lett. B \textbf{299}, 342-344 (1993)
\doi{10.1016/0370-2693(93)90271-I}


%\cite{Frampton:1997in}
\bibitem{Frampton:1997in}
P.~H.~Frampton and M.~Harada,
``Constraints from precision electroweak data on leptoquarks and bileptons,''
Phys. Rev. D \textbf{58}, 095013 (1998)
\doi{10.1103/PhysRevD.58.095013}
[arXiv:hep-ph/9711448 [hep-ph]].




%(ALP)


%\cite{Bauer:2019gfk}
\bibitem{Bauer:2019gfk}
M.~Bauer, M.~Neubert, S.~Renner, M.~Schnubel and A.~Thamm,
``Axionlike Particles, Lepton-Flavor Violation, and a New Explanation of $a_\mu$ and $a_e$,''
Phys. Rev. Lett. \textbf{124}, no.21, 211803 (2020)
\doi{10.1103/PhysRevLett.124.211803}
[arXiv:1908.00008 [hep-ph]].


%\cite{Endo:2020mev}
\bibitem{Endo:2020mev}
M.~Endo, S.~Iguro and T.~Kitahara,
``Probing $e\mu$ flavor-violating ALP at Belle II,''
JHEP \textbf{06}, 040 (2020)
\doi{10.1007/JHEP06(2020)040}
[arXiv:2002.05948 [hep-ph]].


%\cite{Calibbi:2020jvd}
\bibitem{Calibbi:2020jvd}
L.~Calibbi, D.~Redigolo, R.~Ziegler and J.~Zupan,
``Looking forward to lepton-flavor-violating ALPs,''
JHEP \textbf{09}, 173 (2021)
\doi{10.1007/JHEP09(2021)173}
[arXiv:2006.04795 [hep-ph]].


%(muon g-2)

%\cite{Aoyama:2020ynm}
\bibitem{Aoyama:2020ynm}
T.~Aoyama, N.~Asmussen, M.~Benayoun, J.~Bijnens, T.~Blum, M.~Bruno, I.~Caprini, C.~M.~Carloni Calame, M.~C\`e and G.~Colangelo, \textit{et al.}
``The anomalous magnetic moment of the muon in the Standard Model,''
Phys. Rept. \textbf{887}, 1-166 (2020)
\doi{10.1016/j.physrep.2020.07.006}
[arXiv:2006.04822 [hep-ph]].


%\cite{Muong-2:2023cdq}
\bibitem{Muong-2:2023cdq}
D.~P.~Aguillard \textit{et al.} [Muon g-2],
``Measurement of the Positive Muon Anomalous Magnetic Moment to 0.20 ppm,''
[arXiv:2308.06230 [hep-ex]].



% g-2 via LFV mediators

%\cite{Nie:1998dg}
\bibitem{Nie:1998dg}
S.~Nie and M.~Sher,
``The Anomalous magnetic moment of the muon and Higgs mediated flavor changing neutral currents,''
Phys. Rev. D \textbf{58}, 097701 (1998)
\doi{10.1103/PhysRevD.58.097701}
[arXiv:hep-ph/9805376 [hep-ph]].


%\cite{Galon:2016bka}
\bibitem{Galon:2016bka}
I.~Galon, A.~Kwa and P.~Tanedo,
``Lepton-Flavor Violating Mediators,''
JHEP \textbf{03}, 064 (2017)
\doi{10.1007/JHEP03(2017)064}
[arXiv:1610.08060 [hep-ph]].




% electron g-2

%\cite{Parker:2018vye}
\bibitem{Parker:2018vye}
R.~H.~Parker, C.~Yu, W.~Zhong, B.~Estey and H.~M\"uller,
``Measurement of the fine-structure constant as a test of the Standard Model,''
Science \textbf{360}, 191 (2018)
\doi{10.1126/science.aap7706}
[arXiv:1812.04130 [physics.atom-ph]].

%\cite{Morel:2020dww}
\bibitem{Morel:2020dww}
L.~Morel, Z.~Yao, P.~Clad\'e and S.~Guellati-Kh\'elifa,
``Determination of the fine-structure constant with an accuracy of 81 parts per trillion,''
Nature \textbf{588}, no.7836, 61-65 (2020)
\doi{10.1038/s41586-020-2964-7}

%\cite{Fan:2022eto}
\bibitem{Fan:2022eto}
X.~Fan, T.~G.~Myers, B.~A.~D.~Sukra and G.~Gabrielse,
``Measurement of the Electron Magnetic Moment,''
Phys. Rev. Lett. \textbf{130}, no.7, 071801 (2023)
\doi{10.1103/PhysRevLett.130.071801}
[arXiv:2209.13084 [physics.atom-ph]].



%(neutral scalar)

%\cite{Hou:1995dg}
\bibitem{Hou:1995dg}
W.~S.~Hou and G.~G.~Wong,
``$\mu^+ e^- \leftrightarrow \mu^- e^+$ transitions via neutral scalar bosons,''
Phys. Rev. D \textbf{53}, 1537-1541 (1996)
\doi{10.1103/PhysRevD.53.1537}
[arXiv:hep-ph/9504311 [hep-ph]].


%------
%
%(R-parity violating SUSY)


%\cite{Halprin:1993zv}
\bibitem{Halprin:1993zv}
A.~Halprin and A.~Masiero,
``Muonium-antimuonium oscillations and exotic muon decay in broken R-parity SUSY models,''
Phys. Rev. D \textbf{48}, R2987-R2989 (1993)
\doi{10.1103/PhysRevD.48.R2987}




%\cite{Afik:2023vyl}
\bibitem{Afik:2023vyl}
Y.~Afik, P.~S.~Bhupal Dev and A.~Thapa,
``Hints of a new leptophilic Higgs sector?,''
[arXiv:2305.19314 [hep-ph]].

%\cite{Heeck:2023iqc}
\bibitem{Heeck:2023iqc}
J.~Heeck and A.~Thapa,
``Zee-model predictions for lepton flavor violation,''
Phys. Lett. B \textbf{841}, 137910 (2023)
\doi{10.1016/j.physletb.2023.137910}
[arXiv:2303.13383 [hep-ph]].



%\cite{Bahl:2022xzi}
\bibitem{Bahl:2022xzi}
H.~Bahl, J.~Braathen and G.~Weiglein,
``New physics effects on the W-boson mass from a doublet extension of the SM Higgs sector,''
Phys. Lett. B \textbf{833}, 137295 (2022)
\doi{10.1016/j.physletb.2022.137295}
[arXiv:2204.05269 [hep-ph]].

%\cite{Lee:2022gyf}
\bibitem{Lee:2022gyf}
S.~Lee, K.~Cheung, J.~Kim, C.~T.~Lu and J.~Song,
``Status of the two-Higgs-doublet model in light of the CDF $m_W$ measurement,''
Phys. Rev. D \textbf{106}, no.7, 075013 (2022)
\doi{10.1103/PhysRevD.106.075013}
[arXiv:2204.10338 [hep-ph]].



%----------


%\cite{Fukuyama:2022fwi}
\bibitem{Fukuyama:2022fwi}
T.~Fukuyama, Y.~Mimura and Y.~Uesaka,
``Transverse positron polarization in the polarized \ensuremath{\mu^+} decay related with the muonium-to-antimuonium transition,''
Phys. Rev. D \textbf{105}, no.7, 075024 (2022)
\doi{10.1103/PhysRevD.105.075024}
[arXiv:2201.06279 [hep-ph]].



\end{thebibliography}
\end{document}